\documentclass[universe,article,accept,moreauthors,pdftex]{Definitions/mdpi}

\newcommand{\lcdm}{\Lambda\rm{CDM}}

\firstpage{1} 
\pubvolume{1}
\issuenum{1}
\articlenumber{0}
\pubyear{2021}
\copyrightyear{2021}
\externaleditor{Academic Editors: Noemi Frusciante, Francesco Pace and Xue-Mei Deng}
\datereceived{20 November 2021} 
\dateaccepted{16 December 2021} 
\datepublished{18 December 2021} 
\hreflink{https://doi.org/} 

\Title{{Cosmological} 
 Tests of Gravity: A Future Perspective}

\TitleCitation{Cosmological Tests of Gravity: A Future Perspective}

\Author{{Matteo} 
 Martinelli $^{1,2,}${*}
\orcidA{} and Santiago Casas $^{3}$\orcidB{}}

\AuthorNames{Matteo Martinelli, Santiago Casas}

\AuthorCitation{Martinelli, M.; Casas, S.}

\address{%
$^{1}$ \quad Instituto de F\'isica Te\'orica UAM-CSIC, Campus de Cantoblanco, E-28049 Madrid, Spain\\
$^{2}$ \quad INAF---Osservatorio Astronomico di Roma, via Frascati 33, 00040 Monteporzio Catone,  Roma, Italy\\
$^{3}$ \quad Institute for Theoretical Particle Physics and Cosmology (TTK), RWTH
Aachen University, D-52056 Aachen, Germany; casas@physik.rwth-aachen.de}

\corres{Correspondence: matteo.martinelli@inaf.it} 

\abstract{In this review, we outline the expected tests of gravity that will be achieved at cosmological scales in the upcoming decades. We focus mainly on constraints on phenomenologically parameterized deviations from general relativity, which allow to test gravity in a model-independent way, but also review some of the expected constraints obtained with more physically motivated approaches. After reviewing the state-of-the-art for such constraints, we outline the expected improvement that future cosmological surveys will achieve, focusing mainly on future large-scale structures and cosmic microwave background surveys but also looking into novel probes on the nature of gravity. We will also highlight the necessity of overcoming accuracy issues in our theoretical predictions, issues that become relevant due to the expected sensitivity of future experiments.}


\begin{document}

\section{Introduction}

Since its first formulation in 1915, Einstein's general relativity (GR) has demonstrated its ability to pass several observational tests. On~the one hand, Einstein demonstrated that such a theory was able to explain the anomalous precession of Mercury's perihelion, on~the other hand, one of its theoretical predictions, the~gravitational deflection of starlight by the Sun, was tested during the total solar eclipse of 29 May 1919~\cite{Kennefick:2007dr}. Following these early successes, several other predictions of GR passed observational tests in the following century, with~the latest being the detection of gravitational waves~\cite{LIGOScientific:2016aoc} and the observation of a black hole shadow~\cite{EventHorizonTelescope:2019dse}.

From the cosmological point of view, applying GR within the assumption of an expanding, isotropic and almost homogeneous Universe, has allowed to explain most of the observations done by cosmological surveys. However, in~order to satisfy the observational constraints, one has to assume that the Universe is made not only of the particles included in the standard model of particle physics, but~also by two unknown components: cold dark matter (CDM) and dark energy (DE). The~presence of CDM, a~matter component that only interacts gravitationally, is inferred both by cosmological and astrophysical observations (see, e.g.,~in \cite{Corbelli:1999af,Markevitch:2003at,Planck:2018vyg}). These tell us that the amount of standard matter (from here on named baryons) predicted by big bang nucleosynthesis (BBN) cannot account for all the matter content required by observations, and~this unknown CDM component needs to account for $\approx$25\% of the current total energy budget of the Universe. With~the contribution of relativistic particles being negligible at present time, and~baryons accounting for only $\approx$5\% of the total energy, the~remaining $70\%$ of the content of the Universe is made up by DE. Such a component is needed to account for the late time acceleration of the Universe expansion, which was measured for the first time in the late 90s~\cite{SupernovaSearchTeam:1998fmf,SupernovaCosmologyProject:1998vns}. DE is generally identified with the cosmological constant $\Lambda$, whose constant energy density and negative pressure allow to account for this accelerated phase. All these components make up the current standard cosmological model, the~so called cosmological constant-cold dark matter model ($\Lambda$CDM).

Notice that such a model assumes that GR is the correct description of gravity; while GR has been successfully tested, most of the bounds placed on deviations from it come from experiments done within the Solar System, and~the assumption that this theory is a good description of gravity at cosmological scales might still be challenged. In~particular, the~long-known theoretical issues with the cosmological constant (the ``fine tuning'' and ``coincidence'' problems~\cite{RevModPhys.61.1,Velten:2014nra}), together with the recent tensions in the measurements of cosmological parameters (see, e.g., in~\cite{Perivolaropoulos:2021jda}), have prompted the investigation of models alternative to $\Lambda$CDM, including alternative theories of gravity (see, e.g.,~in \cite{DiValentino:2021izs}).

During the last two decades, the~improvement of data from cosmological surveys has made possible to test the very foundation of the $\Lambda$CDM model, i.e.,~the assumption that gravity is described by GR. Cosmic microwave background (CMB) and large-scale structures (LSS) surveys have the ability to probe the way that matter clusters and how it is distributed in the Universe, as~well as giving insight on how matter distorts space-time through gravitational lensing. This allows to constrain possible departures from GR, thus providing observational tests of the theory of gravity at cosmological scales. The~next two decades will see a further improvement of the quality of data, further improving our knowledge of the evolution of the Universe and, consequently, of~the nature of~gravity.

In this review, we want to highlight the progress on this investigation that will be achieved thanks to upcoming cosmological surveys. In~Section~\ref{sec:theory}, we will briefly review the landscape of alternative theories of gravity and we outline the parametric, model independent methods used to constrain departure from GR. We will then outline the currently available cosmological constraints in Section~\ref{sec:currentconstr}. In~Section~\ref{sec:futurecons}, we will review the expected improvement that future surveys will bring on tests of the gravitational theory, also highlighting the role of new cosmological probes. We will also outline in Section~\ref{sec:newchallenges} the new challenges that the improved quality of data will pose, before~summarizing the status and the perspective of this investigation in Section~\ref{sec:summary}.

\section{Testing Gravity at Cosmological~Scales}\label{sec:theory}


In General Relativity it can be shown that the perturbed metric has only two scalar degrees of freedom~\cite{1995ApJ...455....7M} and for the spatially flat Friedmann--Lemaitre--Robertson--Walker (FLRW) spacetime, the~perturbed line element takes the form
\begin{equation}
    ds^2=-(1+2\Psi)dt^2+a(t)^2(1-2{\Phi})dx^i dx_i,
\end{equation} \label{eq:pert-flrw}
where $t$ is the cosmic time, $a(t)$ is the scale factor, and $\Psi(t, x^i)$ and $\Phi(t, x^i)$ are the gravitational scalar potentials. 
In this review, we will concentrate on sub-horizon scales, for~which the wavemodes are $k \gg aH$, meaning that the scales we are interested in are much smaller than the Hubble horizon.
This allows us to work in the quasi-static approximation, in~which the time variation of the gravitational potentials is small compared to the Hubble time and one can neglect the time derivatives of these potentials in the perturbation equations.
This also means that if we were to modify gravity with an extra scalar field, under~this approximation we can also neglect the time derivatives of the scalar field fluctuations below the sound horizon of the scalar perturbation. {While the quasi-static approximation is extremely useful to investigate alternative theories of gravity (see in} \cite{noller2014relativistic, Bellini:2014fua, Pogosian:2016pwr} {for further details), one has to be aware of its limitations; this approximation breaks down for scales larger than the sound horizon of the scalar field in which modifications are encoded, thus making this unsuitable to model observables at very large scales, or~in cases where the speed of sound is much lower than the speed of light} \cite{Sawicki:2015zya, PhysRevD.89.084023}.

When including the full energy-momentum tensor, the~first-order perturbed Einstein equations in Fourier space give two equations that describe the evolution of the two gravitational potentials $\Phi$ and $\Psi$.
These equations read~\cite{caldwell2007constraints, zhao2009searching, amendola2008measuring, hojjati2011mgcamb}
\begin{eqnarray}
-k^{2}\Phi(a,k) & = & 4\pi
Ga^{2}\mu(a,k)\rho(a)\Delta(a,k)\,,\label{eq:mu-def}\\
-k^{2}(\Phi + \Psi)(a,k) & = & 8\pi
Ga^{2}\Sigma(a,k)\rho(a)\Delta(a,k)\,,\label{eq:Sigma-def} 
\end{eqnarray}
where $\rho(a)$ is the average dark matter density and $\Delta(a,k)=\delta+3aH\theta$ is the co-moving density contrast with $\delta$ the fractional overdensity, and~$\theta$ the peculiar velocity. The~so-called Weyl potential $\Psi_{\rm W}$ can be defined as
$
\Psi_{\rm W}(a,k)= (1/2)(\Phi(a,k)+\Psi(a,k))
$
and it is also called the lensing potential, as it describes the propagation of relativistic particles. 
The ratio of the two gravitational potentials is denoted as $\eta$, gravitational anisotropic stress or gravitational slip
\begin{equation}
\eta(a,k) \equiv  \Psi(a,k)/\Phi(a,k)\,\,\,.\label{eq:eta-def}
\end{equation}

The scale- and time-dependent functions $\eta(a,k)$, $\mu(a,k)$ and $\Sigma(a,k)$ stand for all possible deviations of Einstein gravity in these equations, being equal to unity when standard GR is recovered. This approach can encompass any modification by a scalar-tensor theory (see Section~\ref{sec:scaltens}), but~it allows also for more general and even agnostic parameter-free modifications of gravity that do not directly relate to any given model.
Given that there are only two scalar degrees of freedom, it means that of course there is a relationship between $\mu$, $\Sigma$ and $\eta$, and they are related by
\begin{equation}
\Sigma(a,k)=\frac{\mu(a,k)}{2}\left[1+\eta(a,k)\right]\,\,\,.\label{eq:SigmaofMuEta}
\end{equation}

The $\mu(a,k)$ function is usually probed best by experiments that probe the evolution of non-relativistic particles, as these directly trace the evolution of the $\Psi$ potential. The~parameter $\mu$ enters as an additional force affecting tracer particles and therefore it can be cast as an effective Newtonian constant with
\begin{equation}
    \frac{G_{\rm eff}(a,k)}{G} = \mu(a,k)
\end{equation}
in the linear regime and in Fourier space, with~$G$ being the standard Newton constant. 
On the other hand, relativistic particles, and~therefore light, follow the equation for the Weyl potential, meaning that probes of gravitational lensing are mostly sensitive to the function $\Sigma(a,k)$.

For non-relativistic particles and under the assumption of zero anisotropic stress, one can find the growth equation for the density and the velocity perurbations ($\delta$ and $\theta$, respectively) by considering the conservation of the energy-momentum tensor and the linearly perturbed Einstein field equations~\cite{1995ApJ...455....7M}. For~$\lcdm$ in which $\Phi = \Psi$ and in the Newtonian approximation, one can also derive this by looking at the Vlasov--Poisson system of equations~\cite{bernardeau2002large}.
When combining the perturbation equations for $\delta$ and $\theta$, and again considering only sub-horizon scales, one obtains the following differential equation for the growth of matter perturbations ${\delta}_m$:
\begin{equation}
    \ddot{\delta}_m+2{H}\dot {\delta}_m-4 \pi G \bar\rho \mu \delta_m=0, 
\end{equation} \label{eq:deltadotdot}
where the dot represents a derivative with respect to $t$, the~Hubble function is given by $H=\dot{a}/a$ and we can see the effect of deviations from GR  on the growth of perturbations in the $\mu$ function.
Defining the so-called growth rate function via $f = dln\delta/d\ln a$, one can obtain the following equation:
\begin{equation}
    \label{eq:grwthfeq1}
\frac{df}{d \ln a}+f^2+\left(\frac{\dot{H}}{H^2}+2\right)f=\frac{3}{2}\frac{G_{\rm eff}}{G}\Omega_m \equiv \frac{3}{2}\,\mu\, \Omega_m \,.
\end{equation}

This growth rate function $f(k,z)$ enters prominently in the determination of redshift space distortions for the galaxy power spectrum, as~we will see below, and~can be used to constrain redshift- and scale-dependent modifications of gravity that affect the growth of~structures.

While the parameterization of departures from GR through the $\mu$, $\eta$ and $\Sigma$ functions is widely used, there are other ways for these deviations to be described. An~example is another common approach, which we will refer to as $\gamma$-parameterization. Such an approach focuses on possible deviations from the growth of structures expected in GR, encoding these in the growth rate as~\cite{Linder:2005in}
\begin{equation}\label{eq:gammapar}
    f(z) = \left[\Omega_{\rm m}(z)\right]^\gamma\, ,
\end{equation}
where $\gamma$ is a free parameter that in the GR limit reduces to $\gamma\approx0.55$ \cite{1991MNRAS.251..128L,Linder:2005in}.

The two parameterizations can be related to each other, with~the assumption of having modifications in the growth rate leading to $\Sigma=1$, and~the $\mu$ function that can be written, dropping for simplicity the $z$-dependence of $\Omega_{\rm m}(z)$, as~\cite{Mueller:2016kpu}
\begin{equation}
    \mu(a,\gamma) = \frac{2}{3}\Omega_{\rm m}^{\gamma-1}\left[\Omega_{\rm m}^{\gamma}+2+\frac{H'}{H}+\gamma\frac{\Omega'_{\rm m}}{\Omega_{\rm m}}+\gamma'\ln{\Omega_{\rm m}}\right]\, .
\end{equation}

{Another relevant parameterization, often used to constrain deviations from GR, is the $E_G$ statistic} \cite{Zhang:2007nk,Leonard:2015cba,Ghosh:2018ijm}. {This exploit the relation between the $\Phi$ and $\Psi$ potentials in GR and can be used to detect deviations from it using cosmological observables. In~this review, we will not focus further on this particular approach, but~we refer the reader to the the literature that investigates the possible constraints achievable with it (see, e.g.,} in \cite{Ferreira:2019xrr,Blake:2020mzy,Nunes:2021ipq}).

\subsection{Examples of Modified Gravity Models: $f(R)$ and Jordan--Brans--Dicke}

While the functions of Equations~\eqref{eq:mu-def}--\eqref{eq:eta-def} can trace generic deviations from GR, several models alternative to Einstein's theory have been developed, and~their phenomenology can be mapped into these functions. Here, we briefly review how this can be done for two examples of such alternative models: $f(R)$ and Jordan--Brans--Dicke~models.

The set of models known as $f(R)$ theory~\cite{starobinsky1980new}, are obtained when modifying the Einstein--Hilbert action by assuming a function of the Ricci scalar, in~the form $f(R)$ such~that
\begin{equation}
    S = \int \mathrm{d}^4 x \sqrt{-g} f(R) \\
    + 16\pi G \int \mathrm{d}^4 x \sqrt{-g} \mathcal{L}_m(\psi_m^{(i)}, g_{\mu \nu}) \label{eq:fofRaction}
\end{equation}

The field equations can be obtained after a variation of Equation~\eqref{eq:fofRaction} with respect to the metric $g_{\mu \nu}$ and they read
\begin{equation}
    f_R R_{\mu \nu } - \frac{1}{2} f g_{\mu \nu} - \nabla_\mu \nabla_\nu f_R + g_{\mu \nu} \square f_R =  8\pi G T_{\mu \nu} \,\,, \label{eq:fofRfieldeqs}
\end{equation}
where $f_R \equiv \partial f(R) / \partial R$ and $\square$ is the d'Alambertian operator. These equations naturally reduce to Einstein's field equations when $f(R) = R$. 

We can recast the Einstein-frame action of Equation~\eqref{eq:fofRaction} as a scalar field action in the Jordan frame (for a modern review on these frames, see~\cite{catena2007einstein}), by~replacing the $f(R)$ term by $f(\lambda) + (R -\lambda) \mathrm{d}f(\lambda)/\mathrm{d}\lambda$, which is identical to the original $f(R)$ Lagrangian, when varied with respect to the scalar field. 
An auxiliary field $\psi \equiv \mathrm{d}f(\lambda)/\mathrm{d}\lambda$ can be introduced, together with a potential $V(\psi)=m^2_{\rm Pl}(f(\lambda(\psi)) - \lambda(\psi)\psi)/2$. The~scalar field action for $\psi$ is then obtained when replacing back into the original action, and it takes the form
\begin{equation}
    S = \int \mathrm{d}^4 x \sqrt{-g} \left[\psi R - V(\psi) \right] \\
    + 16\pi G \int \mathrm{d}^4 x \sqrt{-\tilde{g}} \mathcal{L}_m(\psi_m^{(i)}, g_{\mu \nu}) \label{eq:jbd-action} \,\,.
\end{equation}

This theory corresponds to the Generalized Jordan--Fierz--Brans--Dicke~\cite{will2014confrontation, clifton2012modified} theory with $\omega_{BD} = 0$.
When this $\omega_{BD}$ parameter is not zero, there is a kinetic term added to the action in the form
$\frac{\omega_{BD}(\psi)}{\psi}\nabla_\mu \psi \nabla^\mu \psi $ and when $\omega_{BD}$ is just a constant, this reduces to the popular Jordan--Brans--Dicke theory~\cite{brans1961mach}.

In the case of $f(R)$ gravity, the~expressions for the gravitational potentials and its modifications defined above in Equations~\eqref{eq:mu-def} and  \eqref{eq:Sigma-def} become relatively simple and reflect the presence of an additional fifth force with a characteristic mass scale~\cite{sawicki2007stability}
\begin{equation}
\label{eq:mass_fR}
m_{f_R}^2\sim \frac{1+f_R}{3f_{RR}}\sim \frac{1}{3f_{RR}}\, ,
\end{equation}
where $f_{RR}$ is the second derivative of the $f(R)$ function with respect to the Ricci scalar $R$.
Assuming negligible matter anisotropic stress and again under the quasistatic approximation, one finds \citep{Pogosian:2016pwr}
\begin{align}
\mu(a,k) &= \frac{1}{1+f_R(a)}\frac {1+4k^2a^{-2}m_{f_R}^{-2}(a) }{1+3k^2a^{-2}m_{f_R}^{-2}(a)}\,,\label{eq:mu_fR}\\
\eta(a,k) &= \frac{1+2{k^2}{a^{-2}m_{f_R}^{-2}(a)}}{1+4k^2a^{-2}m_{f_R}^{-2}(a)}\,, \label{eq:mu_eta_fR}
\end{align}
and
\begin{equation}\label{eq:SigmaHS}
    \Sigma(a)=\frac{1}{1+f_{R}(a)}\,.
\end{equation}

In the case of Jordan--Brans--Dicke, the~modifications $\mu$ and $\eta$ become scale-independent and one can derive again the equations for $\mu$, $\eta$ and $\Sigma$ in the quasi-static approximation~\mbox{\cite{brans1961mach, will2014confrontation, alonso2017observational, joudaki2020testing}}
\begin{align}
    \mu =&\frac{2 (2 + \omega_{\rm BD})}{(3 + 2 \omega_{\rm BD}) \psi}\\
    \Sigma =& \frac{1}{\psi}\\
    \eta =&  \frac{1+\omega_{\rm BD}}{2+\omega_{\rm BD}}  \,.
\end{align}

\subsection{General Scalar-Tensor~Models}\label{sec:scaltens}
The two models described above fall in a more general class of theories, called scalar-tensor models.
For scalar-tensor theories, the action in the Einstein frame, which defines its equations of motion, is expressed as
\begin{equation}
    S = \int \mathrm{d}^4 x \sqrt{-g} \left[\frac{M^2_{Pl}}{2} R -\frac{1}{2}(\nabla \phi )^2 - V(\phi) \right] \\
    + \int \mathrm{d}^4 x \sqrt{-\tilde{g}} \mathcal{L}_m(\psi_m^{(i)}, \tilde{g}_{\mu \nu})  \label{eq:scalarfield-action}
\end{equation}
where $\phi$ is the scalar field, $V(\phi)$ its potential and it couples to the matter fields $\psi_m^{(i)}$ through the Jordan frame metric $\tilde{g}_{\mu \nu}$, which is  related to the metric $g_{\mu \nu}$ as
\begin{equation}
    \tilde{g}_{\mu \nu} = A^2(\phi) g_{\mu \nu}\, . \label{eq:conformal}
\end{equation}

The conformal parameter $A(\phi)$ represents an universal coupling to matter and it implies that particles will feel a total gravitational potential $\Phi_{T}$ which will be the sum of the standard Newtonian term $\Phi_N$ and an additional contribution $\Phi_A$\,,
\begin{equation}
    \Phi = \Phi_N + \Phi_A \,\,,
\end{equation}
whose consequence is the fact that matter particles of mass $m$ are sensitive to a ``fifth force'' given by the gradient of the ``conformal'' contribution to the potential $\Phi_A$.

As can be expected, Equation~\eqref{eq:scalarfield-action} can be generalized to account for all possible theories of a scalar field coupled to matter and the metric tensor. Imposing that the equations of motion are only second order, this general action described the class of Horndeski theories~\cite{horndeski1974second, deffayet2009generalized}.
This general action can be written as
\begin{equation}
    S = \int \mathrm{d}^4 x \sqrt{-g} \left[ \sum^{5}_{i=2} \mathcal{L}_i +  \mathcal{L}_m(\psi_m^{(i)}, g_{\mu \nu}) \right]
\end{equation}
where the four Lagrangian terms corresponds to different combinations of four functions $G_{2,3,4,5}$ of the scalar field and its kinetic energy $\chi=-\partial^\mu \partial_\mu \phi/2 $, the~Ricci scalar and the Einstein tensor $G_{\mu \nu}$ and are given by~\cite{deffayet2011k, kobayashi2011generalized}
\begin{eqnarray}
\nonumber
{\cal L}_{2} & = & K(\phi,\chi),\\ \nonumber
{\cal L}_{3} & = & -G_{3}(\phi,\chi)\Box\phi,\\ \nonumber
{\cal L}_{4} & = & G_{4}(\phi,\chi)\, R+G_{4,\chi}\,[(\Box\phi)^{2}-(\nabla_{\mu}\nabla_{\nu}\phi)\,(\nabla^{\mu}\nabla^{\nu}\phi)]\,,\\ \nonumber
{\cal L}_{5} & = & G_{5}(\phi,\chi)\, G_{\mu\nu}\,(\nabla^{\mu}\nabla^{\nu}\phi) \\ \nonumber
&-&\frac{1}{6}\, G_{5,\chi}\,[(\Box\phi)^{3}-3(\Box\phi)\,(\nabla_{\mu}\nabla_{\nu}\phi)\,(\nabla^{\mu}\nabla^{\nu}\phi) \\
&+&2(\nabla^{\mu}\nabla_{\alpha}\phi)\,(\nabla^{\alpha}\nabla_{\beta}\phi)\,(\nabla^{\beta}\nabla_{\mu}\phi)] \ ,
\label{eq:Horndenski_L}
\end{eqnarray}

After the gravitational wave event GW170817~\cite{LIGOScientific:2017vwq,Goldstein:2017mmi}, which constrained the propagation of gravitational waves to be practically equal to the speed of light~\cite{LIGOScientific:2017zic}, a~large part of Horndeski theory, was ruled out, as~in order to satisfy the observational bound the functions $G_{4,\chi}$, $G_{5,\chi}$ and $G_{5,\phi}$ need to vanish~\cite{baker2017strong}. As~a consequence, the~viable theories within this class remained to be Jordan--Brans--Dicke models and Cubic Galileons (Horndeski theories with Lagrangians up to $\mathcal{L}_3$) \cite{langlois2018scalar, baker2017strong, Ezquiaga:2017ekz, creminelli2017dark, mcmanus2016finding, sakstein2017implications}. See~the work in \cite{ishak2019testing} for a summary on the implications of the GW event on scalar-tensor theories and~beyond.

For a generic Horndeski theory the two functions $\mu$ (expressing the effective gravitational constant) and $\eta$ (the gravitational slip) can be expressed as a combination of five free functions of time $p_{1,2,3,4,5}$, which are related to the free functions $G_i$ in the Horndeski action~\cite{koyama2016cosmological, pogosian2010optimally}
\begin{align}
\mu(a,k) &= \frac{p_1(a) + p_2(a) k^2}{ 1 + p_3(a) k^2}\,,\label{eq:mu_general}\\
\eta(a,k) &= \frac{ 1 + p_3(a) k^2}{ p_4(a) + p_5(a) k^2}\,. \label{eq:eta_general}
\end{align}

{The aim of this review is to focus as much as possible on model independent tests of gravity, but~the Horndeski class of scalar-tensor theories is taken here as an example of a broad class that, while specifying to specif mechanisms of departure for GR, allows to remain reasonably agnostic. However, while the scalar-tensor theories that fall under the Horndeski class cover quite a large part of the alternatives theories of gravity of current interest for cosmology, the~space of currently available theories is significantly wider. Several reviews focusing more in detail on the different classes of modified gravity models are available in the literature (see, e.g.,} in \cite{clifton2012modified,Nojiri:2017ncd,CANTATA:2021ktz, koyama2016cosmological,ishak2019testing}{), and~we refer the reader to these should they be interested in this topic.}

\subsection{The $\alpha$-Parametrization in Modified~Gravity} \label{sec:alpha}

A physically meaningful parametrization of the linear Horndeski action, given bythe work in ~\cite{ Bellini:2014fua}, is related to the Effective Field Theory of dark energy~\cite{Gubitosi:2012hu,Bloomfield:2012ff,Piazza:2013coa}, where all possible deviations to the scalar-tensor action are parameterized linearly. This parametrization  is of  great help when discussing current cosmological constraints.
It is defined using  four functions of time---$\alpha_M$, $\alpha_K$, $\alpha_B$ and $\alpha_T$---plus the effective Planck mass $M_\star^2$ and a function of time for a given  background specified by the time variation of the Hubble rate  $H(a)$ as a function of the scale factor $a$.
The term $\alpha_T$ measures the excess of speed of gravitational waves compared to light and after the event GW170817, this term is constrained to be effectively zero. The~term
$\alpha_K$ quantifies the kineticity of the scalar field and therefore appears in models like K-mouflage, which require the  K-mouflage screening~\cite{Brax:2015pka} in order to pass solar system constraints. The~coefficient
$\alpha_B$ quantifies the braiding or mixing of the kinetic terms of the scalar field and the metric and can cause dark energy clustering. It appears in modified gravity models where a fifth force is present~\cite{Pogosian:2016pwr}.
Finally, $\alpha_M$ quantifies the running rate of the effective Planck mass and it is generated by a non-minimal coupling. This parameter modifies the lensing terms, as it directly affects the lensing potential~\cite{ishak2019testing}.

While this $\alpha$-parameterization is not easily mapped into the $\mu$, $\eta$ and $\Sigma$ functions (see~in \cite{ Bellini:2014fua} for specific mappings under certain limits), and~it refers to the specific class of Horndeski theories rather than to generic deviations from GR, obtaining observational constraints on the $\alpha_i$ functions is extremely useful to investigate the theory of gravity, as~it allows to directly relate possible deviations from the expected GR behavior to physical~effects.

\subsection{Impact on Cosmological~Observables}

In Equations~\eqref{eq:mu-def} and  \eqref{eq:Sigma-def}, we saw how the functions parameterizing departures from GR enter in the Poisson equations for relativistic and non-relativistic particles, while we have shown explicitly in Equation~\eqref{eq:grwthfeq1} how the $\mu$ function enters the expression for the growth rate of cosmological perturbations. These effects naturally enter the equations used to obtain theoretical predictions on observables, thus imprinting signatures of the gravity theory that can be detected by cosmological~surveys.

Given their impact on the growth of structures and on gravitational lensing, natural candidates to test gravity are the LSS probes of galaxy clustering (GC) and  weak lensing (WL).
GC measures the two-point correlation function of galaxy positions either in three dimensions, i.e.,~angular positions and redshift, or~using a two-dimensional tomographic approach (angular galaxy clustering) when the redshift information is not particularly good. The~first approach is used for spectroscopic surveys, which provide a very small error on the redshift of the observed galaxies, while the second is more suited for photometric~surveys.

For the former, one works in Fourier space, where the correlation function of galaxies, known as the observed galaxy power spectrum $P^{obs}_{gg}$ is directly related to the power spectrum of matter density perturbations $P_{\delta \delta, zs}$ in redshift space by~\cite{ballinger1996measuring, baldauf2015equivalence, wang2010designing, taruya2010baryon}
\begin{equation}
P^{\mathrm{obs}}_{\rm gg}(z,k,\mu_{\theta})=  {\rm AP}(z) P_{\delta \delta, {\rm zs}}(k, z) E_{\rm err}(z,k)  + P_{\rm shot}(z) \,,\label{eq:P_obs}
\end{equation}
where $AP(z)$ corresponds to the Alcock--Paczynski effect, $E_{err}(z,k)$ is a damping term given by redshift errors  and $P_{\rm shot}(z)$ is the shot noise from estimating a continuous distribution out of a discrete set of points. $\mu_{\theta}$ is the cosine of the angle between the line of sight and the wave vector $\mathbf{k}$.
Furthermore, the~redshift space power spectrum, is given by~\cite{wang2010designing, taruya2010baryon}
\begin{equation}
P_{\delta \delta, {\rm zs}}(z,k,\mu_{\theta})=  {\rm FoG}(z, k, \mu_{\theta}) K^{2}(z, \mu_{\theta} ; b(z); f(z)) P_{\delta \delta} (k,\mu_{\theta},z) \,,\label{eq:P_zs}
\end{equation}
 where ${\rm FoG}(z, k, \mu_{\theta})$ is the ``Fingers of God'' term that accounts for nonlinear peculiar velocity dispersions of the galaxies and K is the redshift space distortion term that depends---in linear theory, where it is known as the Kaiser term~\cite{kaiser1987clustering}---on the growth rate $f(z)$ and the bias $b(z)$, but~can be more complicated when taking into account nonlinear perturbation theory at mildly nonlinear scales~\cite{scoccimarro1995loop, carrasco2012effective, bernardeau2002large}. For~a detailed explanation of these terms, we refer the reader to the work in ~\cite{Euclid:2019clj} and the many references~therein.

As it can be seen in Equation~\eqref{eq:P_zs}, the~growth rate enters the $K$ term, thus imprinting in it possible signatures of departures from GR through Equation~\eqref{eq:grwthfeq1}. In~addition to this, also the $P_{\delta\delta}$ contains information on the $\mu$ function, as~it is obtained through Equation~\eqref{eq:mu-def}, thus making a spectroscopic survey of GC an ideal probe to test~gravity.

While not able to provide a 3D information, a photometric GC survey would be able to provide constrain on deviations from the standard clustering of matter. In~this case, one compares the angular power spectrum of galaxy correlations, obtained in tomographic redshift bins, with~the theoretical predictions obtained as~\cite{Euclid:2019clj}
\begin{equation}\label{eq:cijdefgc}
C_{ij}^{gg}(\ell) = \frac{c}{H_0} 
\int{\frac{{\hat{W}}_i^g(z) {\hat{W}}_{j}^g(z)}{E(z) r^2(z)}
	P_{\delta\delta}\left ( k_{\ell}, z \right ) dz}\,,
\end{equation}
where the $\hat{W}_i^g$ functions contain information on the galaxy distribution in the $i$-th bin and the galaxy bias needed to relate the correlation of galaxies to that of matter, $r(z)$ is the co-moving distance and $E(z)=H(z)/H_0$ is the dimensionless Hubble parameter. Even though Equation~\eqref{eq:cijdefgc} does not depend on the $K$ term containing the growth rate, deviations from GR would still affect such an observable through its dependence on $P_{\delta\delta}$.

Photometric galaxy surveys also allow to measure the distortion of the shape of distant galaxies caused by the lensing effect produced by the matter distribution between them and the observers. The~measurement of such a shear does not rely on the measurement of galaxy correlation functions and it is therefore a direct tracer of matter distribution, avoiding the dependence on the galaxy bias $b(z)$. Furthermore, for this observable, one can obtain the angular power spectrum~\cite{Euclid:2019clj}
\begin{equation}\label{eq:cijdef}
C_{ij}^{\gamma\gamma}(\ell) = \frac{c}{H_0} 
\int{\frac{{\hat{W}}_i^\gamma(z) {\hat{W}}_{j}^\gamma(z)}{E(z) r^2(z)}
	P_{\Phi+\Psi}\left ( k_{\ell}, z \right ) dz}\,,
\end{equation}
where the window functions, or~lensing kernels, $\hat{W}_{j}^\gamma(z)$ contain information on the geometry of space-time, and~$P_{\Phi+\Psi}\left ( k_{\ell}, z \right )$ is the Weyl power spectrum, coming from the Weyl potential defined Equation~\eqref{eq:Sigma-def}. This is related to the matter power spectrum $P_{\delta \delta}$ by
\begin{equation}\label{eq:weylLAM}
P_{\Phi+\Psi} =  \Sigma^2(k,z) \left[3\left(\frac{H_0}{c}\right)^2\Omega_\mathrm{m}^{0} (1 + z)\right]^2 P_{{\delta \delta}}\,.
\end{equation}

In this equation, we can see clearly the observational signature of the $\Sigma$ lensing function defined above in Equation~\eqref{eq:Sigma-def}, thus making WL of distant galaxies an ideal probe to constrain deviations from the lensing effect expected in GR. Notice that in Equation~\eqref{eq:cijdef} we neglected the contribution of intrinsic alignment, a~systematic effect for such a measurement that is described in detail in~\cite{Euclid:2019clj}.

The discussion above highlights the potential of LSS observations to test the theory of gravity, as~the combination of GC and WL can bring information on two of the MG functions, $\mu$ and $\Sigma$.

\begin{figure}[H]
	
	\begin{tabular}{cc}
	\includegraphics[width=0.45\columnwidth]{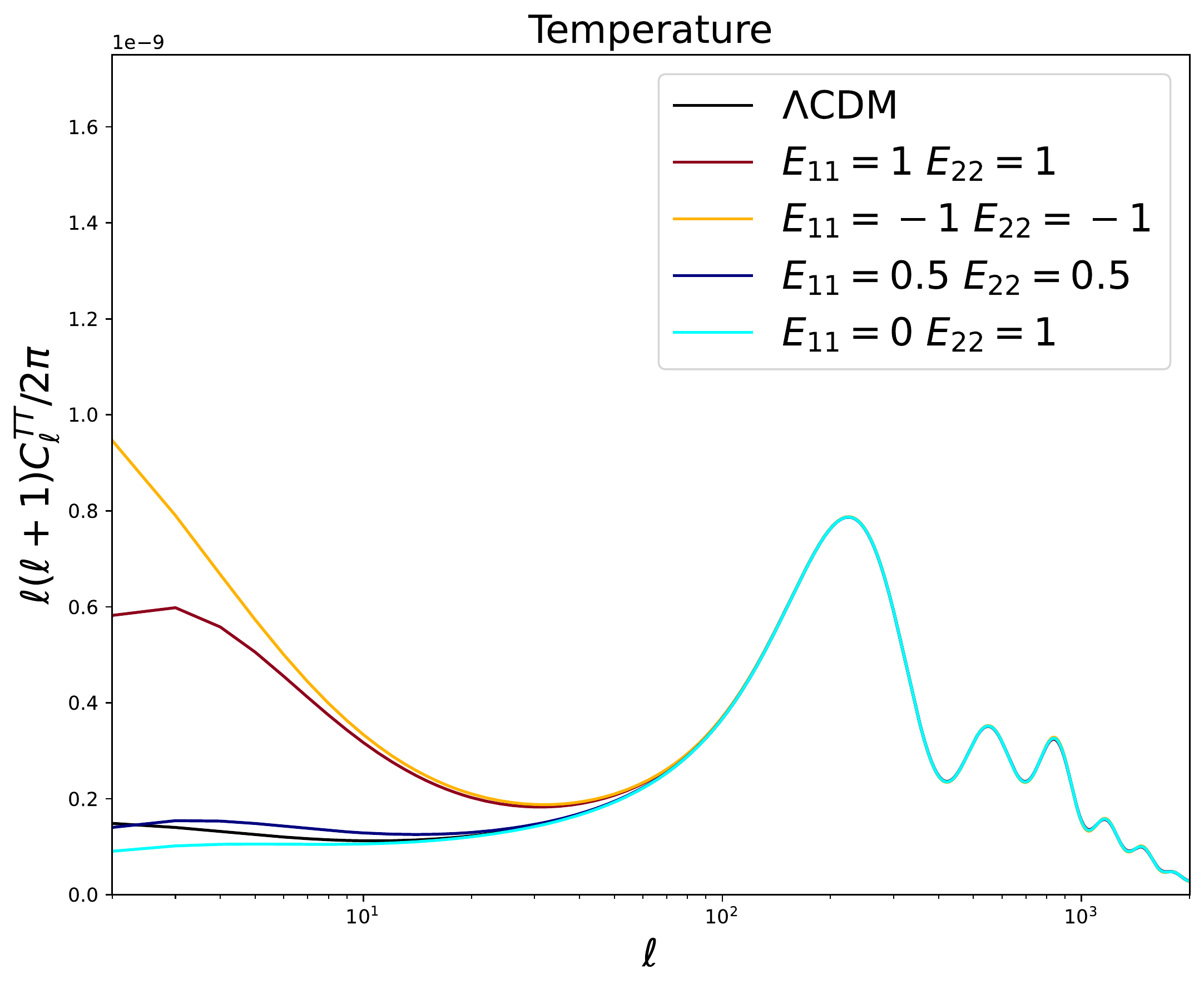} &
	\includegraphics[width=0.45\columnwidth]{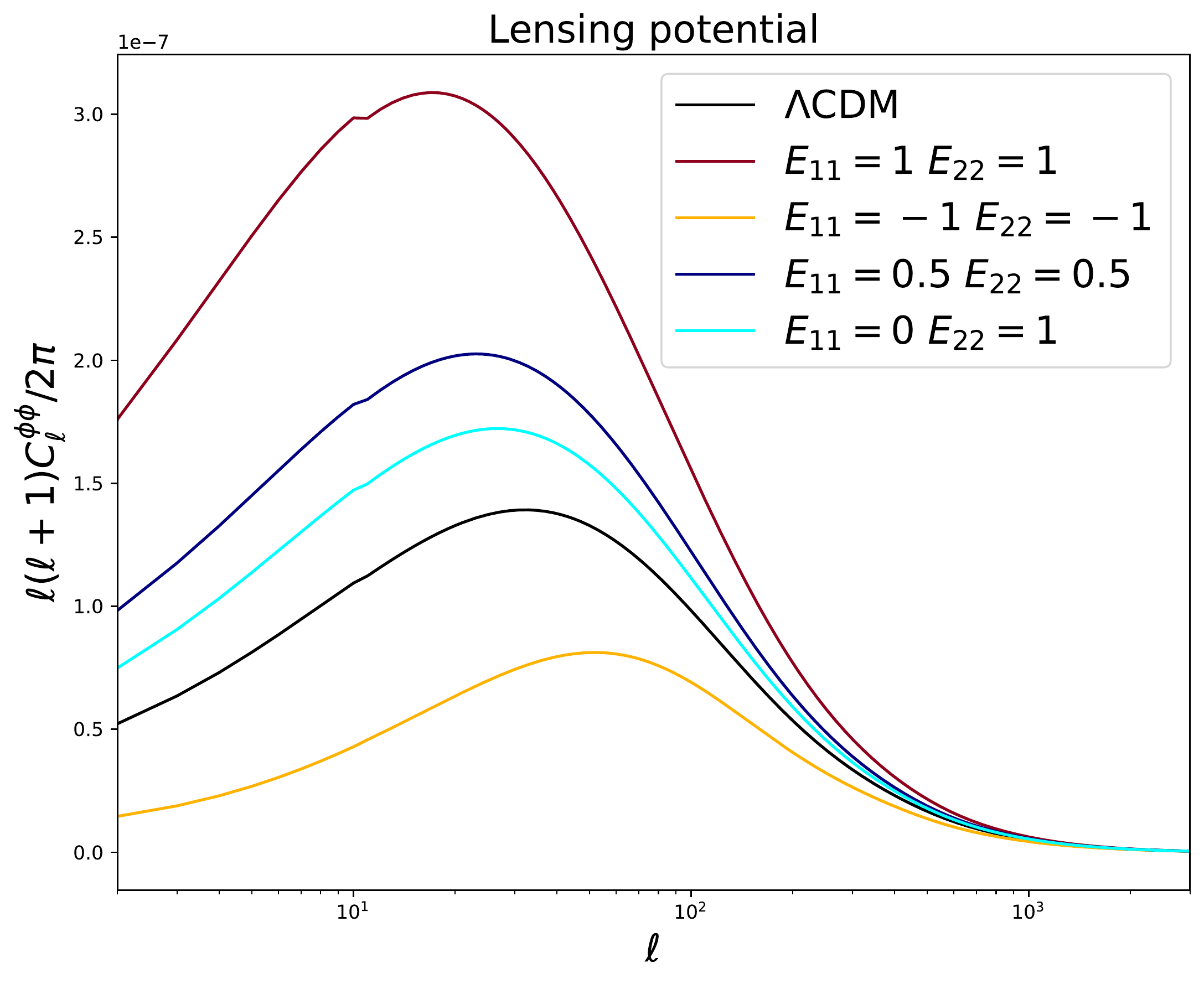} \\
	\end{tabular}
	\caption{Temperature (\textbf{left}) and lensing potential (\textbf{right}) CMB power spectra obtained for possible departures from GR. This figure is obtained using the same parameterization and parameter values in~\cite{Planck:2015bue} using a private~Boltzmann-solver.}\label{fig:cmbspectra}
\end{figure}

Alongside LSS observables, another mean to test the theory of gravity is through observations of the cosmic microwave background (CMB). This radiation is composed of photons coming from very high redshift, and~the anisotropies in their temperature and polarization are related to the small perturbations in the matter--photon plasma in the early Universe. While departures from GR are expected to be relevant only at late times, signatures of these can still be detected or ruled out using CMB observations. The~anisotropies in the CMB angular power spectra are indeed affected also by a number of secondary effects cause by the distribution of matter and its effect on photons during their travel to the observer from the last scattering surface. The~main signatures of the theory of gravity on CMB spectra can indeed be found in the integrated Sachs--Wolfe (ISW) effect~\cite{1967ApJ...147...73S,1985SvAL...11..271K}, caused by the time evolution of the potentials $\Phi$ and $\Psi$, and~in the lensing effect on CMB peaks~\cite{Acquaviva:2005xz,Carbone:2013dna}, which is related to the Weyl potential. We show in Figure~\ref{fig:cmbspectra} an example of the effects of deviations from GR on CMB temperature and lensing potential power spectra, following the approach of the Planck Collaboration~\cite{Planck:2015bue}, highlighting the sensitivity of CMB measurements to signatures of the gravity theory. Such predictions were obtained using Equations~\eqref{eq:mu-def} and \eqref{eq:eta-def}, with~the $\mu$ and $\eta$ functions parameterized as
\begin{equation}\label{eq:mupk}
\mu(z) = 1+E_{11}\Omega_{\rm DE}(z)\,,
\end{equation}
\begin{equation}\label{eq:etapk}
\eta(z) = 1+E_{22}\Omega_{\rm DE}(z)\,,
\end{equation}
where $E_{ii}$ are free parameters and $\Omega_{\rm DE}(z)$ is the abundance of the dark energy~component.

{Note that at this point, while we have focused in this section on perturbation observables, and~we will do so throughout the paper, tests of gravity can be performed also using observations of the background expansion of the Universe. Indeed, models alternative to GR usually predict an expansion history that departs from the one expected in $\Lambda$CDM, and~observations of this can be used to constrain such models. Furthermore, such an investigation can be performed in a model independent way, e.g.,~by employing a cosmographic approach} \cite{Sahni:2002fz,Alam:2003sc}{, which would allow to exploit to use observations of supernovae and baryon acoustic oscillations to constrain deviations from a cosmological constant driven Universe, and~to connect the results to possible modifications of gravity (see, e.g.,} in \cite{Capozziello:2019cav}). {Noteworthy, other observables could even allow to reconstruct the expansion history at even higher redshift, e.g.,~through Quasars or $\gamma$-ray bursts (see, e.g.,} in \cite{Capozziello:2008tc,Rezaei:2020lfy,Bargiacchi:2021hdp}).

\subsection{Codes and Tools to Compute Cosmological~Observables}

Theoretical predictions for the evolution of cosmological perturbations and, consequently, for~cosmological observables, can be obtained very efficiently when these perturbations are small, and~linear perturbation theory can be applied. The~so-called Einstein--Boltzmann solvers apply this formalism to evolve primordial perturbations to present time and to obtain prediction on the observables that we described ago. Two commonly used public codes of this kind are  \texttt{CAMB}\endnote{\url{https://camb.info}, {accessed}  on 15 December 2021 
} (Code for Anisotropies in the Microwave Background) \cite{Lewis:1999bs,2012JCAP...04..027H}, which is written mainly in \texttt{fortran}, and~\texttt{CLASS}\endnote{\url{https://lesgourg.github.io/class_public/class.html}, {accessed} on 15 December 2021} (Cosmic Linear Anisotropy Solving System) \cite{lesgourgues2011cosmic, blas2011cosmic}, which is mainly written in the \texttt{C} programming language. Both of these codes come with user-friendly \texttt{python} wrappers.
These codes can be modified to account for alternative theories of gravity, or~to account for parameterized departures from GR. While in the first case once implements specific modified gravity (MG) models or generic class of models solving their full scalar field equations, in~the latter one modifies the evolution equations through Equations~\eqref{eq:mu-def} and \eqref{eq:eta-def}.

The most commonly used codes implementing the generic parametrization approach are \texttt{ISitGR}\endnote{\url{https://labs.utdallas.edu/mishak/isitgr/}, {accessed}  on 15 December 2021} \cite{Dosset2011, Dossett2012}, \texttt{MGCAMB}\endnote{\url{https://github.com/sfu-cosmo/MGCAMB}, {accessed}  on 15 December 2021} \cite{hojjati2011mgcamb, zucca2019mgcamb} and, more recently, a~branch of \texttt{CLASS}, called \texttt{QSA\_CLASS} \cite{pace2021comparison}.
For the implementation of specific classes of alternatives to GR we will mention here the two most important ones, namely, \texttt{hi\_class}\endnote{\url{http://miguelzuma.github.io/hi\_class\_public/}, {accessed}  on 15 December 2021} \cite{zumalacarregui2017hi_class} and \texttt{EFTCAMB}\endnote{\url{http://eftcamb.org/}, {accessed}  on 15 December 2021} \cite{hu2014eftcamb, Hu:2014oga}. The~first one, works directly on the Horndeski action of Equation~\eqref{eq:Horndenski_L} and its five free functions of time or, alternatively, by~using the $\alpha$-parametrization defined in Section~\ref{sec:alpha}. On~the other hand, \texttt{EFTCAMB} works within the formalism of the Effective Field Theory of Dark Energy~\cite{Gubitosi:2012hu,Bloomfield:2012ff,Piazza:2013coa}, even though it allows a reparameterization of the EFT functions by means of the $\alpha_i$.

\section{Current Constraints on Modified~Gravity}\label{sec:currentconstr}
The current state of the art for constraints on deviations from GR is given mainly by CMB and LSS~observations. 

Current CMB constraints come mainly from the observations of {\it Planck} \cite{Planck:2018nkj}, a~satellite survey that obtained data from 2009 to 2013; the Planck Collaboration released in 2015 a paper dedicated to constraints on DE and modified gravity models, containing also bounds on possible deviations from GR~\cite{Planck:2015bue}, which were then updated in 2018 with the last data release of the collaboration~\cite{Planck:2018vyg}. 

\begin{figure}[H]
	
	\begin{tabular}{cc}
	\includegraphics[width=0.45\columnwidth]{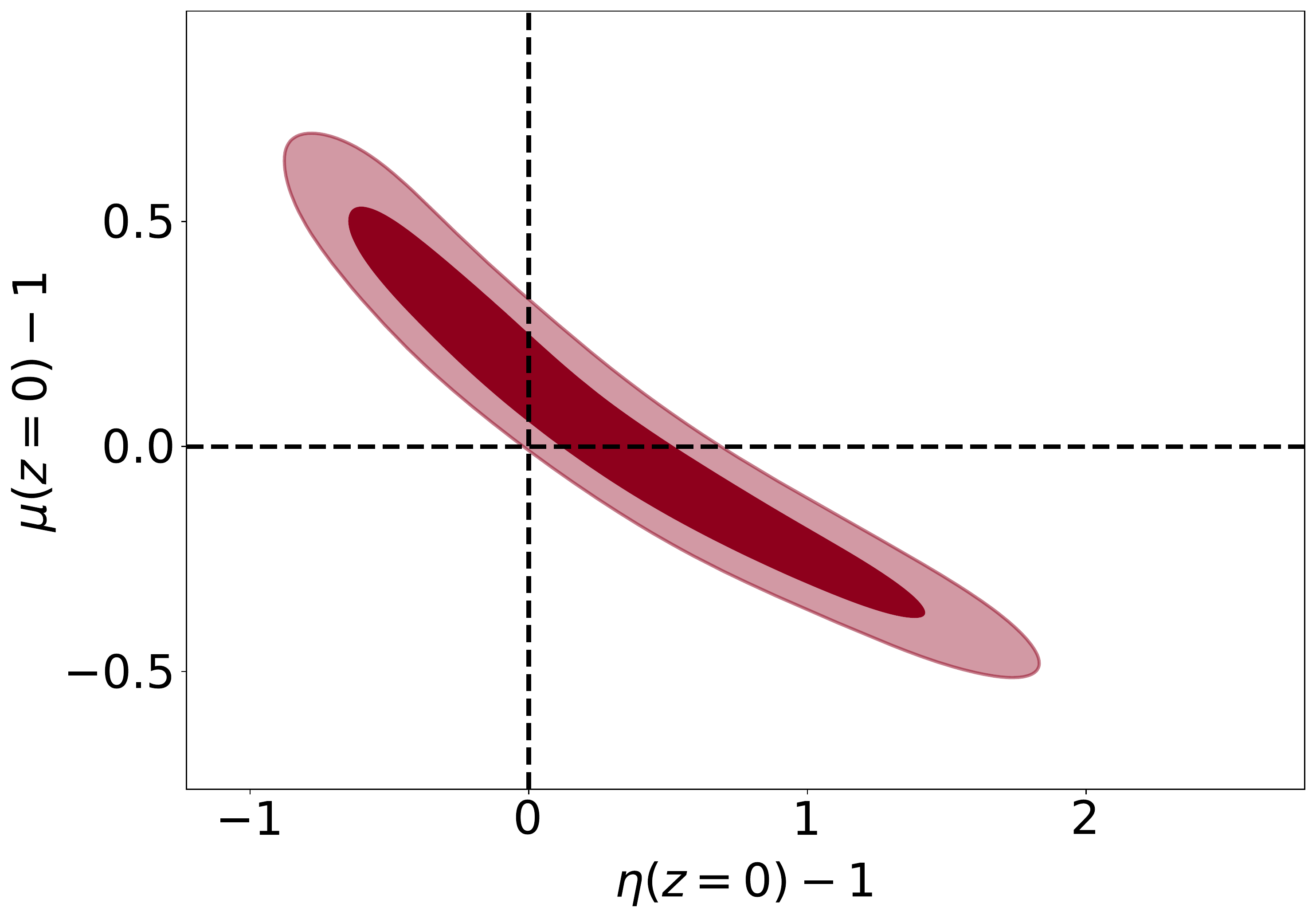} &
	\includegraphics[width=0.45\columnwidth]{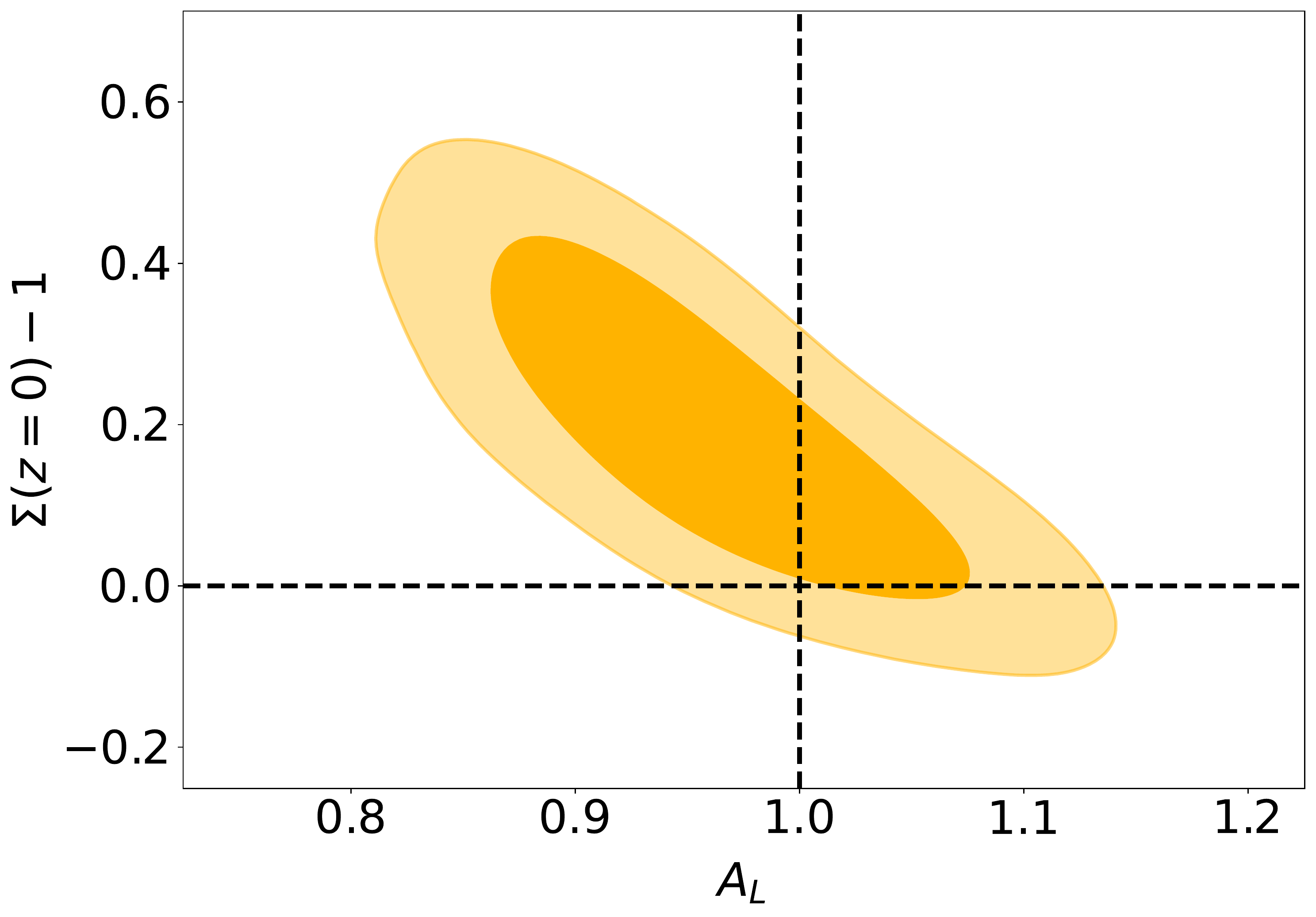} \\
	\end{tabular}
	\caption{{$68\%$} 
 and $95\%$ confidence level contours for $\mu(z=0)-1$ and $\eta(z=0)-1$ (\textbf{left}), and~for $\Sigma(z=0)-1$ and $A_L$ (\textbf{right}). The~dashed lines show the expected values in a GR-$\Lambda$CDM~Universe.}\label{fig:planck}
\end{figure}

In their latest results, the~collaboration constrained deviations from GR parameterized as in Equations~\eqref{eq:mupk} and ~\eqref{eq:etapk}. We show in Figure~\ref{fig:planck} the results obtained fitting the {\it Planck} data using the parameterization of Equations~\eqref{eq:mupk} and \eqref{eq:etapk} through \texttt{MGCAMB}, corresponding to those reported in~\cite{Planck:2018vyg}. One can notice how the $\mu(z)$ and $\eta(z)$ function are compatible with their GR limit within $2\sigma$. As~pointed out in~\cite{Planck:2015bue,Planck:2018vyg}, this agreement is worsened if CMB lensing reconstruction data are removed from the data combination; this is due to the fact that {\it Planck} data prefer a lensing amplitude higher than what one would expect in standard $\Lambda$CDM, a~preference that is reduced when lensing extraction data are included, and~that can be explained by a departure from GR. The~right panel of Figure~\ref{fig:planck} shows this effect highlighting how the $\Sigma(z)$ function is degenerate with $A_L$, the~phenomenological parameter used to take into account deviations from a standard lensing effect~\cite{Calabrese:2008rt}.

Following the results of {\it Planck}, the~observables useful to improve constraints on deviations from GR were those related to LSS observations. In~particular, the~{Dark Energy Survey (DES)} collaboration, after~its latest data release, has employed a subset of its data, the~DMASS galaxy catalog~\cite{DES:2019ikh}, in~combination with {baryon acoustic oscillations (BAO) and redshift space distortions (RSD)} data from CMASS~\cite{BOSS:2016goe}, and~the CMB data from {\it Planck} to further improve the bounds on possible deviations from GR. 
The analysis performed uses a slightly different parameterization from the one of Equations~\eqref{eq:mupk} and \eqref{eq:etapk}, relying instead on~\cite{DES:2021zdr}
\begin{equation}
    \mu(z)=1+\mu_0\frac{\Omega_{\rm DE}(z)}{\Omega_{\rm DE}^0},
\end{equation}
\begin{equation}
    \Sigma(z)=1+\Sigma_0\frac{\Omega_{\rm DE}(z)}{\Omega_{\rm DE}^0},
\end{equation}
where $\Omega_{\rm DE}^0$ is the present time abundance of dark energy, and~the GR limit is achieved for $\mu_0=\Sigma_0=0$. The~results obtained show a slight deviation from GR when DMASS and CMASS are combined, with~$\mu_0=-1.23^{+0.81}_{-0.83}$ and $\Sigma_0=-0.17^{+0.16}_{-0.15}$. However, when {\it Planck} data are included in the analysis, the~GR limit is consistent with the results, with $\mu_0=0.37^{+0.47}_{-0.45}$ and $\Sigma_0=0.078^{+0.078}_{-0.082}$ \cite{DES:2021zdr}.

Other results are available thanks to current data (see, e.g., in~\cite{Joudaki:2016kym,DES:2018ufa}), but~qualitatively the same conclusions can be taken; the GR limit of the parameterized departures used is in agreement with the observational data or has a tension that is not statistically significant \mbox{($\lesssim$2$\sigma$)}. However, the~bounds obtained would still allow for some of the available alternative gravity models to be viable and to be candidates to explain the accelerated expansion of the Universe. In~order to further put GR to test and possibly to rule out some of its alternatives, the~next generation of observational surveys will be~crucial.

\section{Upcoming~Constraints}\label{sec:futurecons}

In the previous section we summarized the current constraints on deviations from GR achievable with currently available data. Here, we focus instead on how Einstein's theory could be tested in the upcoming~decades.

We expect stronger constraints thanks to the improvement in sensitivity that future LSS and CMB surveys will have. This means that the same approach used until now can be applied using future surveys to constrain GR, and~we discuss the expected improvement in Sections~\ref{sec:lssforecast} and \ref{sec:cmbforecast}. 

However, new cosmological probes are now reaching a level of maturity that could allow us to use them to test gravity, and~we will review the constraining power that these will bring to the investigation in Section~\ref{sec:newprobes}.

\subsection{LSS~Forecast}\label{sec:lssforecast}

Next-generation galaxy surveys will be one of the most powerful probes of cosmology in the upcoming decade. These were designed to measure cosmological parameters with $1\%$ precision, through the use of two-point statistics likeGC and WL.
Among the most important upcoming missions there will be {the Javalambre-Physics of the Accelerated Universe Astrophysical Survey (J-PAS) \cite{J-PAS:2014hgg}, the~Dark Energy Spectroscopic Instrument (\textsc{DESI})} \cite{collaboration2018desi}, {\it Euclid} \cite{amendola2018cosmology, laureijs2011euclid}, the~Vera Rubin Observatory~\cite{2019ApJ...873..111I}, WFIRST~\cite{2015arXiv150303757S} and the SKA Observatory~\cite{SKA:2018ckk}.


{\it Euclid} is a European space satellite mission that has an infrared spectrograph and a visible camera. The~latter is capable of taking spectra of about 50 million galaxies in the sky from redshifts $0.9 < z < 1.6$ with a sub-percent accuracy in the redshift estimation, perfect for spectroscopic galaxy clustering studies. On~the other hand, the~visible camera will be able to take low-noise level pictures of \textasciitilde1 billion galaxy shapes and positions up to a redshift of about $z\lesssim 3.0$, which makes it adequate to detect the tiny cosmic shear effect statistically, by~looking at the change of ellipticities of the galaxies across space and time. It will cover an approximate area of 15,000 deg$^2$ in the sky~\cite{Euclid:2019clj, laureijs2011euclid}. {\it Euclid} is expected to be launched by late~2022. 

WFIRST, a~NASA mission, will have similar capabilities in terms of area coverage, instruments and redshift range, but~it will also serve other purposes, such as the search for exoplanets and detailed studies of stars~\cite{spergel2013wfirst}.
On the spectroscopic side, but~based on the ground, \textsc{DESI} will be at the forefront of the observation capabilities in the next years, with~its mission starting in 2021 and with very good spectroscopic resolution, it will be able to capture up to 50 million galaxies on the night sky. It will be able to differentiate between several populations of galaxies, such as Luminous Red Galaxies (LRGs), Emission Line Galaxies (ELGs) and it also will be able to measure quasars and the Lyman-$\alpha$ forest, which will allow the determination of independent clustering statistics along the history of the Universe~\cite{levi2013desi}.
On the side of photometric observables, the~Vera Rubin Observatory will be the main ground-based experiment in the upcoming decade. Its Legacy Survey of Space and Time (LSST) will cover an area of about 20,000 deg$^2$ and it will be capable of obtaining accurate shapes and angular positions of about  a billion galaxies up to redshift $z \lesssim 3.0$. It will also serve other purposes such as the study of Supernovae and transients, making it a very versatile observatory~\cite{abell2009lsst}. {Another photometric survey relevant for the purpose of gravity tests is J-PAS, a~wide field survey carried out from the Javalambre Observatory in Spain, which started observations in 2015, and~which will observe an area of 8500 deg$^2$.}

The SKA Obesrvatory (SKAO) will instead provide a different window for cosmological investigation, as~it will perform observations in the radio band. It will be the largest radio array ever built and it will provide observations of galaxy clustering and weak lensing from resolved galaxies and intensity mapping, through radio continuum and 21-cm line emission~\cite{SKA:2018ckk}.

LSS surveys are extremely interesting for the purpose of testing gravity; these will be able to map the distribution of matter in the Universe by probing its evolution from primordial perturbations and the gravitational lensing effect that it has on photons (see Section~\ref{sec:theory}). For~this reason, the~literature on this topic is extensive, and~we report here a subset of the forecast results available as an~example.

Most of the available forecast results, rely on the Fisher matrix approach~\cite{vogeley1996eigenmode, tegmark1997measuring, tegmark1997karhunen}, that allows to forecast the constraining power of future experiments. Despite its limitations, like assuming a Gaussian likelihood and a Gaussian posterior, the~Fisher matrix $\mathbf{F}$ allows us to obtain an optimistic bound~\cite{carron2013assumption} on the confidence contours of the model parameters, given a prior and a set of experimental settings.
$\mathbf{F}$ can be obtained by knowing the theoretical model (in the case of LSS, the~observable power spectra) and its derivatives with respect to the parameters, together with a model for the experimental noise. The~individual uncertainties on the cosmological parameters $\theta_i$ are then obtained by taking the inverse of $\mathbf{F}$, the~so-called forecasted covariance matrix $\mathbf{C}$, with~ $\mathbf{C}=\mathbf{F}^{-1}$ and looking at its diagonal elements, so that $\sigma(\theta_i) = \sqrt{C_{ii}}$.

\begin{figure}[H]
	
	\begin{tabular}{cc}
	     \includegraphics[width=0.45\columnwidth]{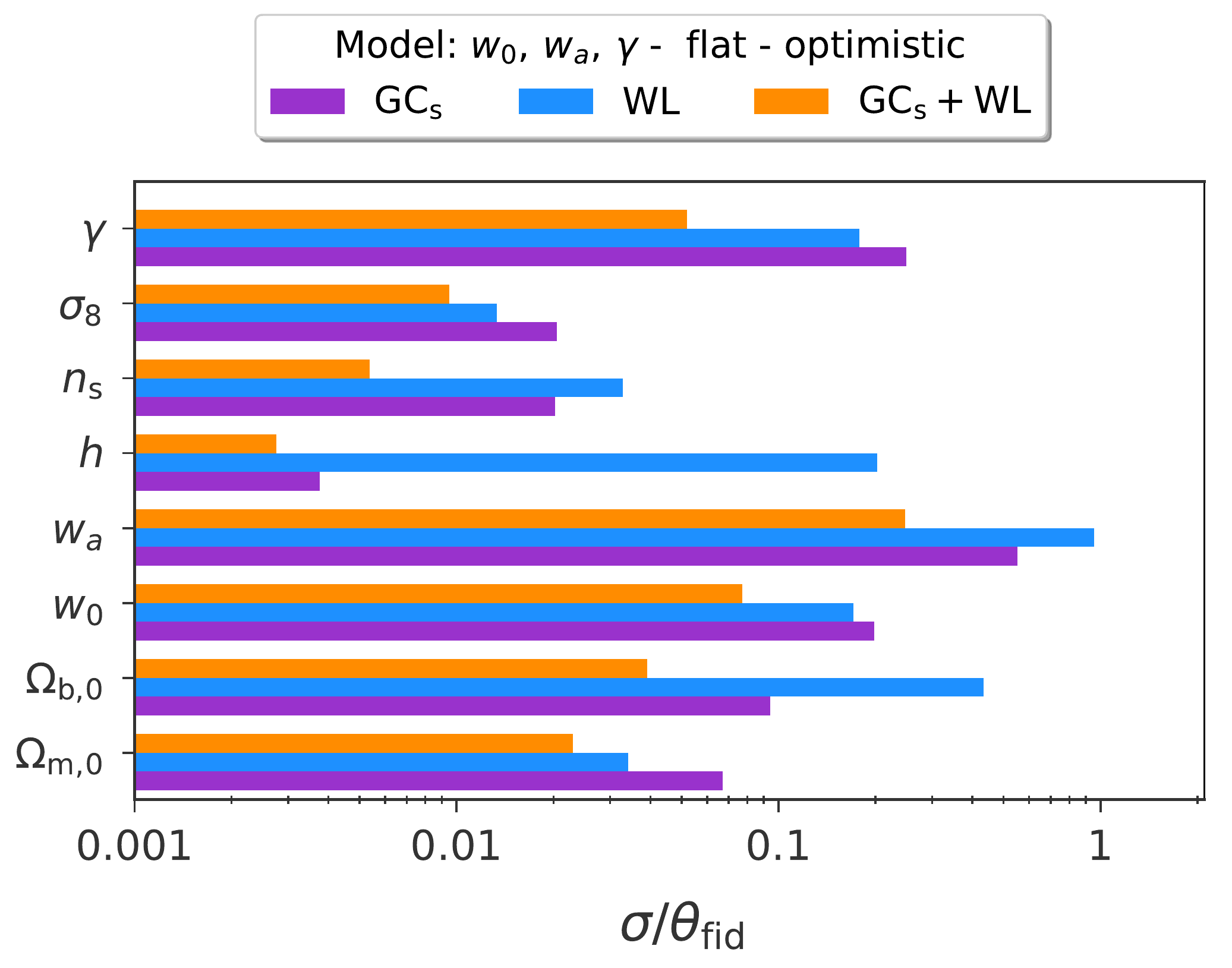}&  
	     \includegraphics[width=0.45\columnwidth]{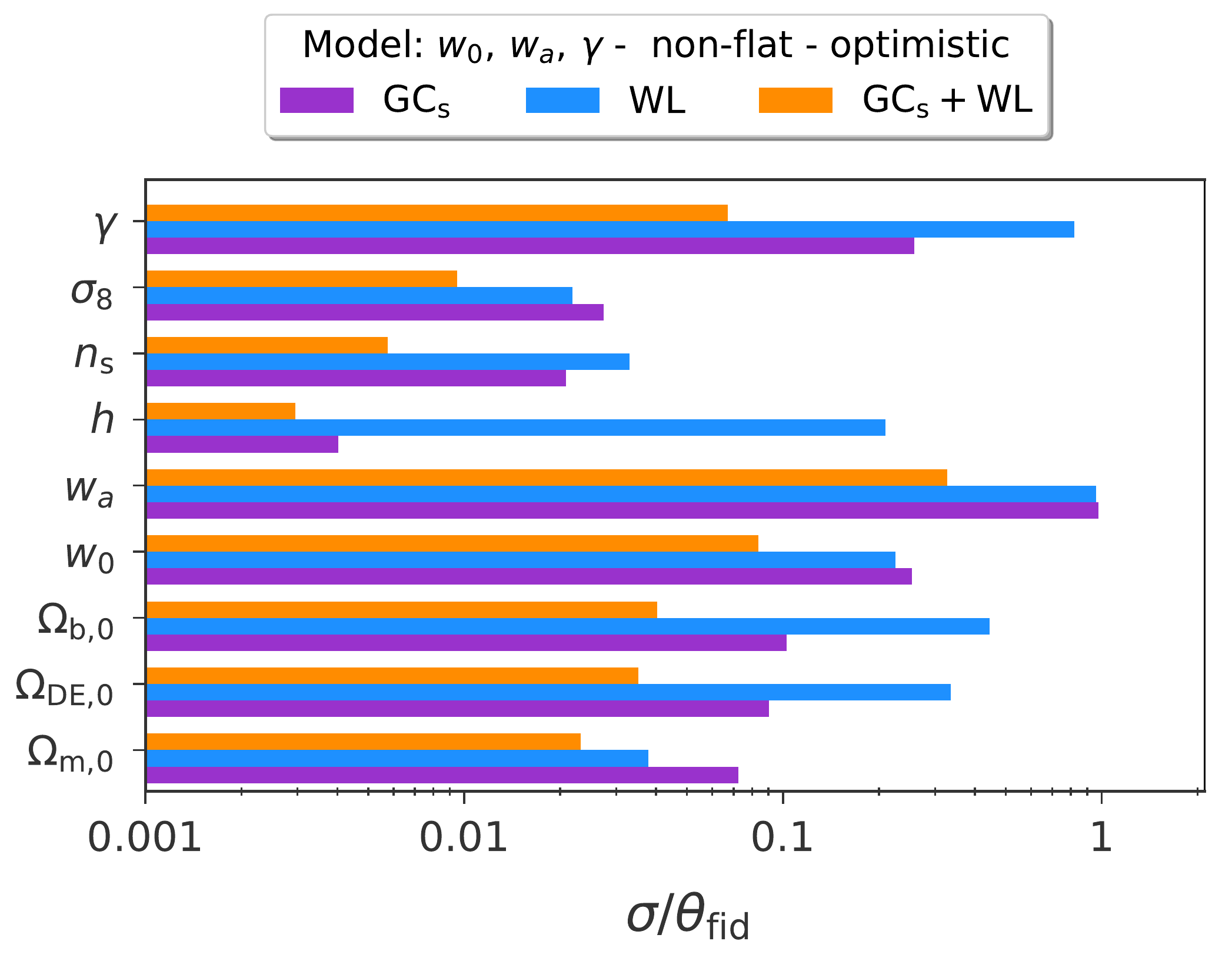}
	\end{tabular}
	\caption{Relative errors on cosmological parameters and on the $\gamma$ controlling the deviation from the standard growth. The~results shown here include spectroscopic galaxy clustering (purple), weak lensing (blue) and their combination (orange). This figure is obtained in \cite{Euclid:2019clj}.}\label{fig:euclidgamma}
\end{figure}

Such a method has been applied with several parameterizations for deviations from GR. Constraints on the $\gamma$ parameter of Equation~\eqref{eq:gammapar} have been obtained by the Euclid Consortium~\cite{Euclid:2019clj}, whose results are shown in Figure~\ref{fig:euclidgamma}. The~authors found that combining photometric and spectroscopic observables (WL and spectroscopic GC), the~$\gamma$ parameter can be measured with a relative error $\lesssim$0.1, thus tightly constraining deviations from the growth of structure expected in GR, even in the case where the background expansion is given freedom to deviate from the standard cosmological constant, and~the assumption of a flat Universe is~dropped.

Forecast constraints from {\it Euclid} have been also obtained for the functions $\mu(z)$ and $\Sigma(z)$ that parameterize deviations from both the standard growth and lensing effect. The~authors of~\cite{Casas:2017eob} constrained these functions exploring two different assumptions: a ``late-time'' parameterization where the two functions scale with the DE energy density $\Omega_{\rm DE}$ and an ``early-time'' parameterization where the redshift trend of $\mu(z)$ and $\Sigma(z)$ is described with a Taylor expansion. Moreover, the~authors explored a case in which the two functions are binned in redshift, an~approach that allows to avoid any assumption on the evolution of these functions, at~the expense of a higher number of parameters and, therefore, looser constraints. This analysis was performed assuming values of the free parameters that do not coincide with the GR limit, but~are still not ruled out by {\it Planck} measurements (see Section~\ref{sec:currentconstr}).

\begin{figure}[H]
	
	\includegraphics[width=0.65\columnwidth]{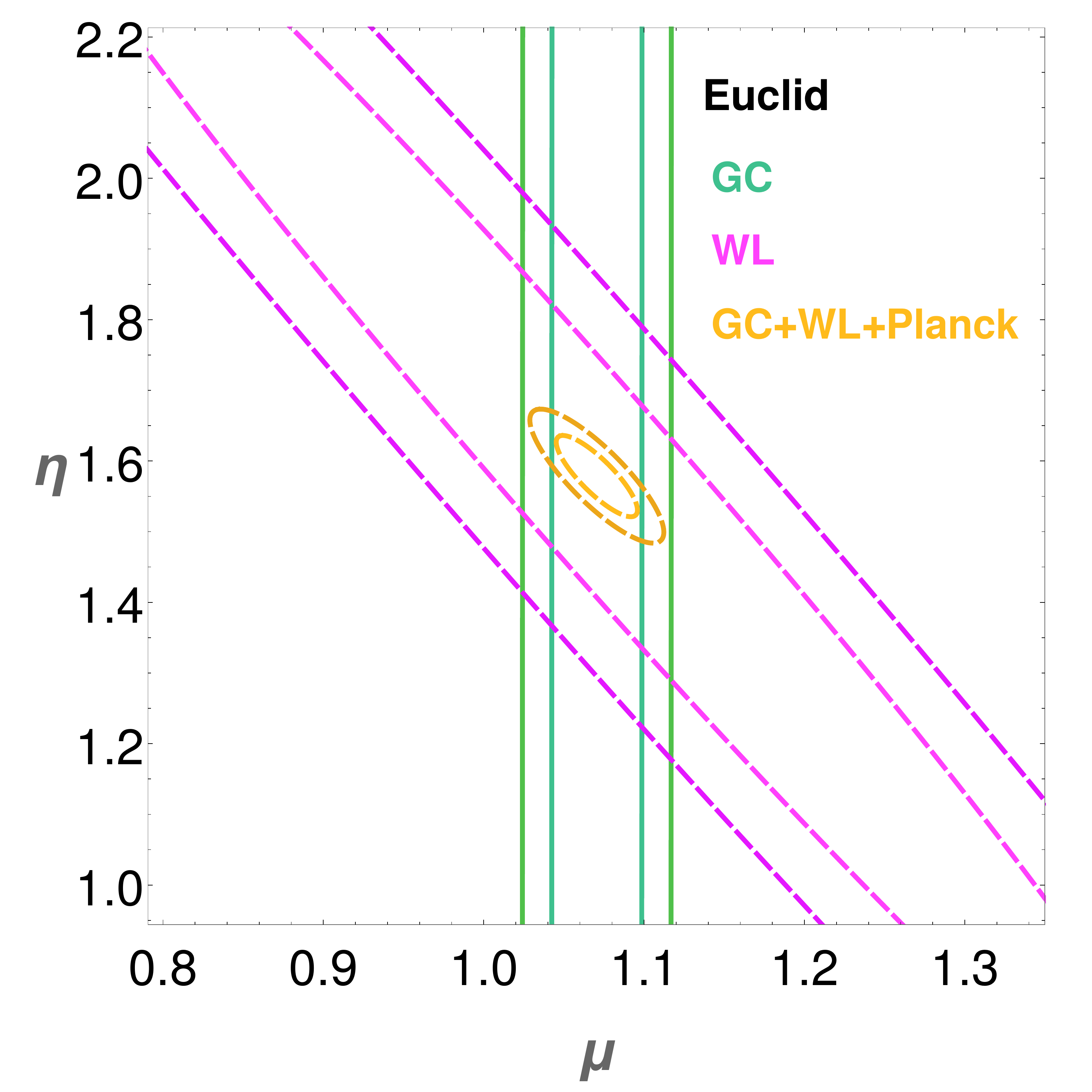}
	\caption{$68\%$ and $95\%$ confidence level contours on the current values of $\mu(z)$ and $\Sigma(z)$ obtained using forecast observations for {\it Euclid}'s spectroscopic GC (green) and photometric WL (purple). The~combination of the two is studied with CMB data from {\it Planck} also included (orange). Reprinted from~the work in \cite{Casas:2017eob}, with~permission from~Elsevier.}\label{fig:euclidmusigma}
\end{figure}

In Figure~\ref{fig:euclidmusigma}, we show the constraints obtained in~\cite{Casas:2017eob} when the 'late-time' parameterization is assumed, i.e.,
\begin{equation}\label{eq:latemu}
    \mu(z) = 1+E_{11}\Omega_{\rm DE}(z)\, ,
\end{equation}
\begin{equation}\label{eq:lateeta}
    \eta(z) = 1+E_{22}\Omega_{\rm DE}(z)\, ,
\end{equation}
where $E_{11}$ and $E_{22}$ are the free parameters ruling the amplitude of the deviation from GR, and~the two functions can be combined to obtain $\Sigma(z)$ following Equation~\eqref{eq:SigmaofMuEta}. These results highlight how the GC and WL probes are complementary, with~the former measuring the clustering of matter (and therefore constraining the $\mu(z)$ function), while the latter is more sensitive to the $\Sigma(z)$ function, as~it probes the shear of galaxy shapes due to gravitational lensing. Combining these two probes, {\it Euclid} can reach constraints of $1.6\%$ and $1\%$ on the present day values of $\mu(z)$ and $\Sigma(z)$ respectively in the late-time parameterization, with~the precision reaching $0.7\%$ and $0.6\%$ if instead the measurements for the two probes are assumed to come from the planned phase 2 of SKAO. In~the binned case, the~best constrained value of $\mu(z)$ is measured with a precision of $2.2\%$, while for $\eta(z)$ the best precision in $3.6\%$. Such a result highlights how upcoming surveys will allow to test gravity even without a priori assumptions on the redshift trend of the parameterized deviations from~GR.

Notice that the constraints of~\cite{Casas:2017eob} are obtained with a specific prescriptions for the nonlinear evolution of cosmological perturbations, which allows to include small scales in the analysis (see Section~\ref{sec:newchallenges}).

{Constraints on a parameterization analogous to the one of} Equations~\eqref{eq:latemu} and \eqref{eq:lateeta} {were also obtained for the expected observations of J-PAS} \cite{AparicioResco:2019hgh}. {The authors of this paper show how the constraints achievable on the parameters ruling the amplitude of deviations from GR are competitive with those that can obtained from {\it Euclid}, thanks to the amount of data that J-PAS will obtain at low redshift, where such parameterization departs the most from the standard model. In~this paper, the~authors also explored the possibility for this functions to be scale-dependent, highlighting how future surveys such as J-PAS and {\it Euclid} can potentially detect a deviation from the scale independent behavior of GR.}

The binned approach to constrain the MG functions $\mu(z)$ and $\eta(z)$ was also used in other forecast for upcoming missions (see, e.g., in~\cite{Asaba:2013xql,Hojjati:2013xqa}). Of~particular interest is the approach of~\cite{Hojjati:2013xqa}, where the binning of the functions, together with a principal component analysis (PCA) \cite{goodfellow2016machine}, was used to constrain these functions when the assumption that they are scale independent is dropped. The~authors explored both the case in which the dependence on the scale $k$ is arbitrary, thus binning the functions in both $z$ and $k$, and~the case in which the $k$-dependence is obtained by requiring that $\mu(z,k)$ and $\eta(z,k)$ come from physically motivated theories~\cite{Silvestri:2013ne}. For~this latter case, the~authors write the MG functions as in Equations~\eqref{eq:mu_general} and \eqref{eq:eta_general}, and~they bin in redshift the $p_i$ functions. The~results in \cite{Hojjati:2013xqa} show that the LSST survey, while unable to constrain individual $p_i$ functions due to their degeneracies, will be able to measure combinations of all of them with a precision of $1\%$ on the best constrained eigenmode, a~measurement which will allow to constrain generic deviations from~GR.

As discussed in Section~\ref{sec:theory}, an~alternative to the $\mu(z)$ and $\Sigma(z)$ function to test gravity is the use of the $\alpha_i$ functions of Section~\ref{sec:alpha}. Constraints on these functions, that consider only departures from GR that fall under the Horndeski class, were also forecasted for upcoming LSS~surveys.

In~\cite{alonso2017observational}, the authors explore the possibility to constrain the $\alpha_i$ functions parameterized~as
\begin{equation}
    \alpha_i = c_i \frac{\Omega_{\rm DE}(z)}{\Omega_{\rm DE}(z=0)}\,.
\end{equation}

\begin{figure}[H]
	
	\includegraphics[width=0.65\columnwidth]{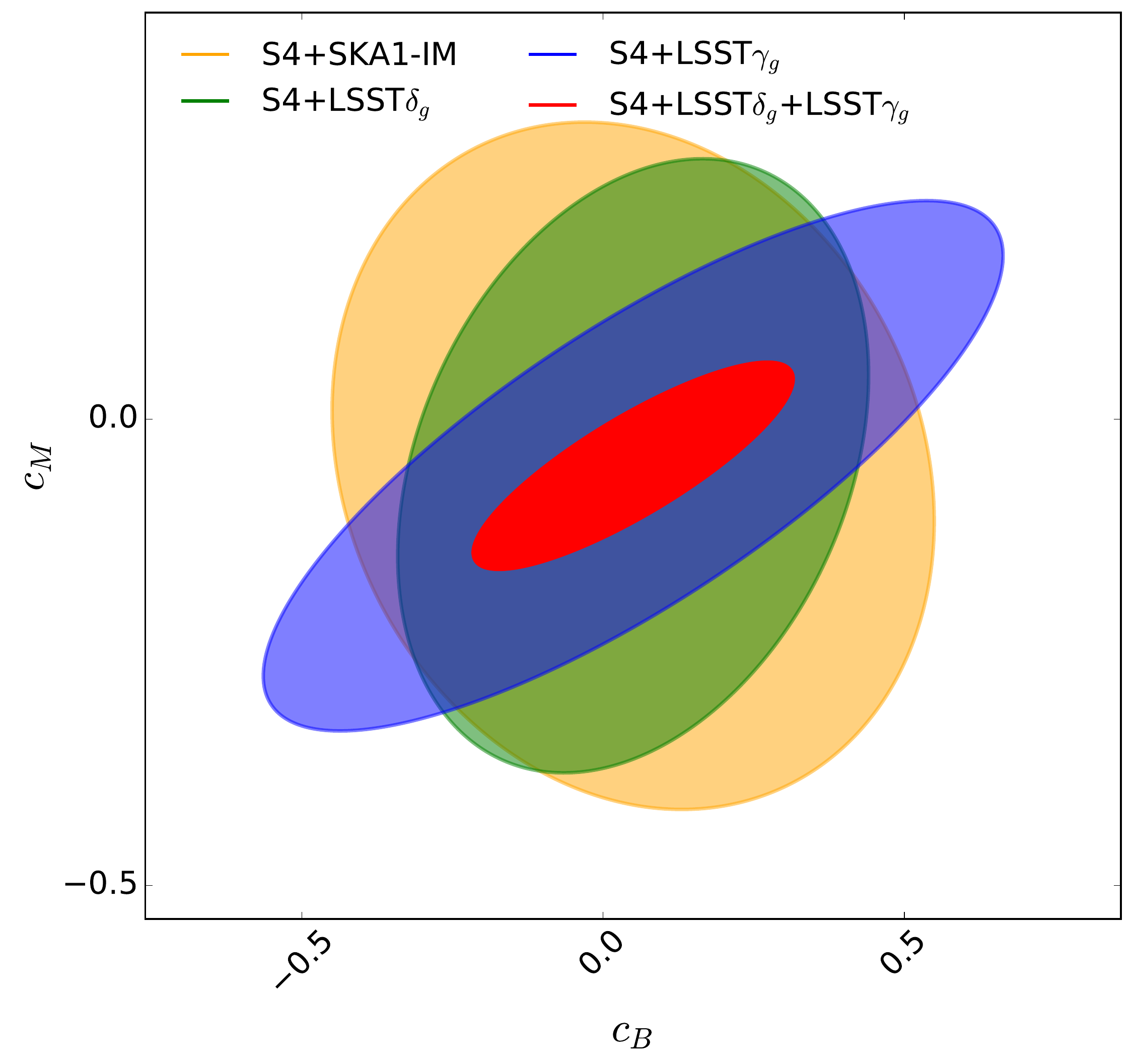}
	\caption{Sixty-eight percent (68\%) confidence level contours on $c_M$ and $c_B$ obtained combining a stage IV CMB survey with the expected measurements of intensity mapping from SKAO (yellow), weak lensing (blue) and galaxy clustering (green) from LSST, and~the combination of the latter two (red). Reprinted figure with permission from the authors of~\cite{alonso2017observational} by the American Physical~Society.}\label{fig:alphasalonso}
\end{figure}

They explore the bounds that can be obtained on the $c_i$ parameters from LSST and also using the intensity mapping survey of SKAO. We show in Figure~\ref{fig:alphasalonso} an example of the results of this analysis, showing the complementarity of the different probes. In~particular one can notice how the combination of LSST clustering and weak lensing observations can significantly tighten the constraints on $c_M$ and $c_B$, ruling the deviation from the standard Planck mass and braiding term~respectively.

\begin{figure}[H]
	
	\includegraphics[width=0.85\columnwidth]{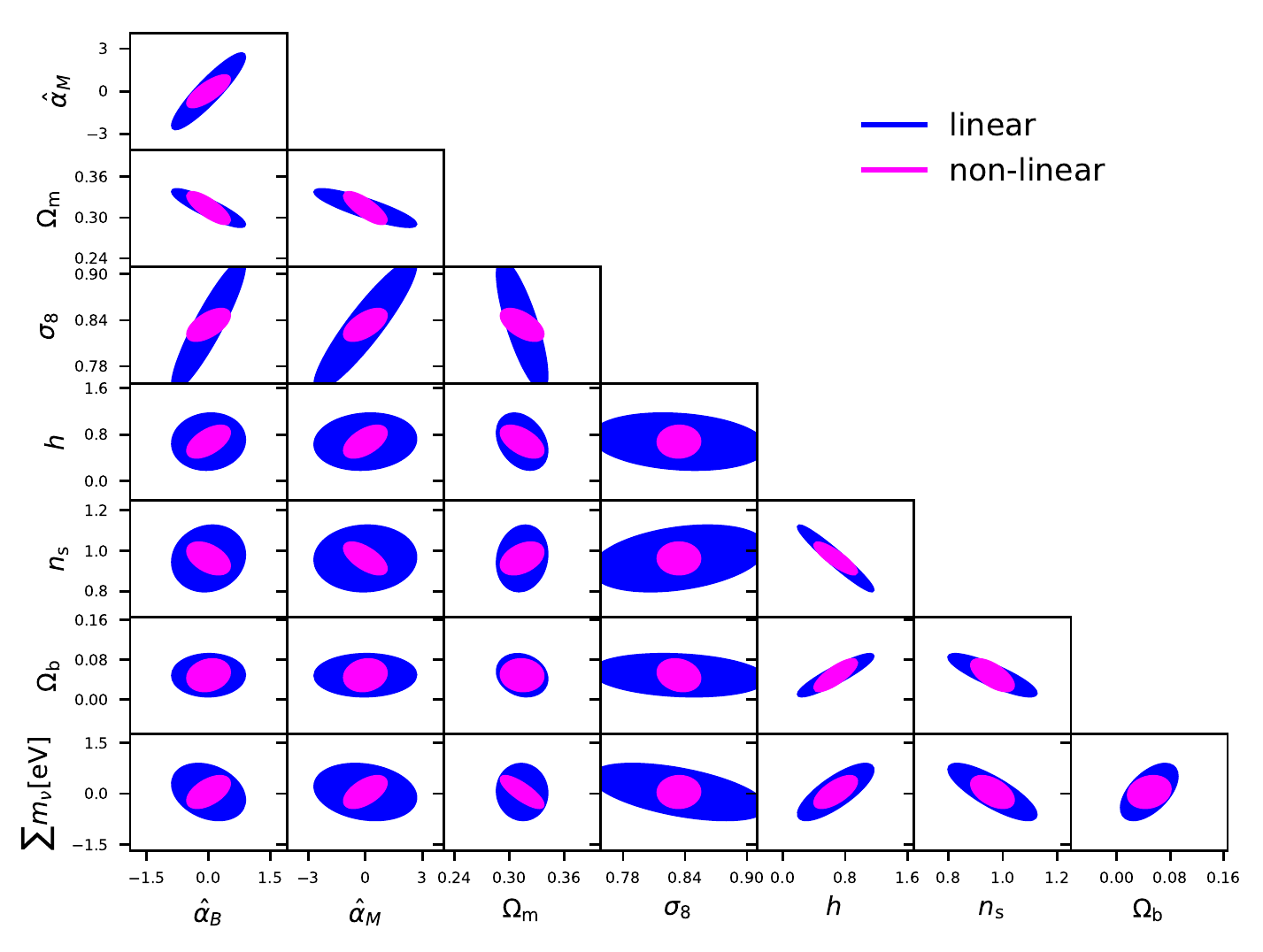}
	\caption{$68\%$ confidence level contours on cosmological and MG parameters obtained by a {\it Euclid}-like survey cutting the nonlinear scales out of the analysis (blue) or including them (pink). This figure is taken from~\cite{spurio2018testing}.}\label{fig:nonlinearalpha}
\end{figure}

Another analysis, performed in~\cite{spurio2018testing}, focuses on the impact of the choice of analysis method on the expected bounds on the $\alpha_i$, comparing the tomographic approach for weak lensing to an analysis that fully considers 3D information, finding that the latter can improve the constraints of $\approx$20\%. Moreover, they investigate they impact of cutting the small scales out of the analysis; such a cut is necessary when one is not able to obtain theoretical predictions for scales at which the linear perturbation description is not valid anymore, which is a common issue for models that deviate from $\Lambda$CDM. As~shown in Figure~\ref{fig:nonlinearalpha} the impact of nonlinear scales can be critical to obtain stringent test of gravity, but~the to fully exploit the data in such a regime can be extremely challenging, as~we will discuss in Section~\ref{sec:newchallenges}.

We conclude this section by stressing that in realistic settings, these constraints might suffer from degeneracies with nuisance parameters that account for observational effects such as intrinsic alignment of galaxies, galaxy bias and other systematic effects, which, if~not accounted for, can severely limit the robustness of the constraints obtained on deviations from GR~\cite{laszlo2012disentangling,ferte2019testing}.

\subsection{CMB} \label{sec:cmbforecast}

While the current CMB data provided by {\it Planck} have reached the cosmic variance limit for a wide range of multipoles in temperature, i.e.,~the correlation where most on the information on gravity can be extracted, we still expect an improvement in the constraints on deviations from GR from upcoming CMB~surveys.

On the one hand, upcoming surveys and especially those performed by ground based telescopes, will be able to access smaller scales and therefore access higher multipoles with respect to {\it Planck}. On~the other hand, upcoming surveys will significantly improve the sensitivity for polarization spectra, for~which the cosmic variance limited data provided by {\it Planck} is much more limited in the range of multipoles. Both these improvement will allows to extract more information on the CMB lensing potential, by~measuring temperature power spectra oscillation at high multipoles, and~by improving the lensing reconstruction methods that also rely on a precise measure of polarization spectra~\cite{Planck:2018lbu}.

As discussed in Section~\ref{sec:currentconstr}, most of the CMB constraining power on gravity comes from the impact of lensing on CMB observables; thus, thanks to the future improvements in its characterization, we can expect better constraints on deviations from~GR.

For the purpose of testing gravity, the~main CMB surveys expected in the upcoming decades are the Simons Observatory (SO) \cite{SimonsObservatory:2018koc} and CMB-S4~\cite{CMB-S4:2016ple}. The~former is a new CMB experiment that is being built in Chile, while the second is an envisioned CMB survey that would consist of dedicated facilities at the South Pole, Chile and possibly a northern hemisphere~site.

Most of the forecast obtained for future CMB surveys have been obtained in the context of scalar-tensor theories, rather than as generic deviations from GR, thus aiming to constrain either the $G_i$ functions included in the Horndeski Lagrangian, e.g.,~through the parameterization done with the $\alpha$ functions described in Section~\ref{sec:alpha}.

Constraints on this kind of theories are indeed one of the science cases proposed in the CMB-S4 science book~\cite{CMB-S4:2016ple}; the authors considered simulated observations for this proposed experiment and~use {\it Planck} measurements to complete the sky map in the areas not surveyed by CMB-S4 to obtain a full sky map. They find that the inclusion of CMB-S4 data can improve the constraints on the parameters ruling the deviations from GR in scalar-tensor theories, in~particular with the bound on the speed of gravitational waves parameter $\alpha_T$ that is improved by a factor $\approx$3, a~constraint mainly due to the increased sensitivity of such a survey on primordial gravitational~waves.

Despite the fact that future CMB surveys will improve our constraints on deviations from GR, such surveys will not be competitive by themselves with contemporary experiments mapping the LSS. The~latter kind of surveys is indeed more suited to map the distribution of matter in the Universe and the evolution of perturbations at times when one expects deviations from GR to be relevant (if present). However, both SO and CMB-S4 surveys will contribute to this investigation when used in combination with LSS surveys, as~they will break degeneracies between cosmological and MG parameters~\cite{SimonsObservatory:2018koc,CMB-S4:2016ple}.

\begin{figure}[H]
	
	\includegraphics[width=0.85\columnwidth]{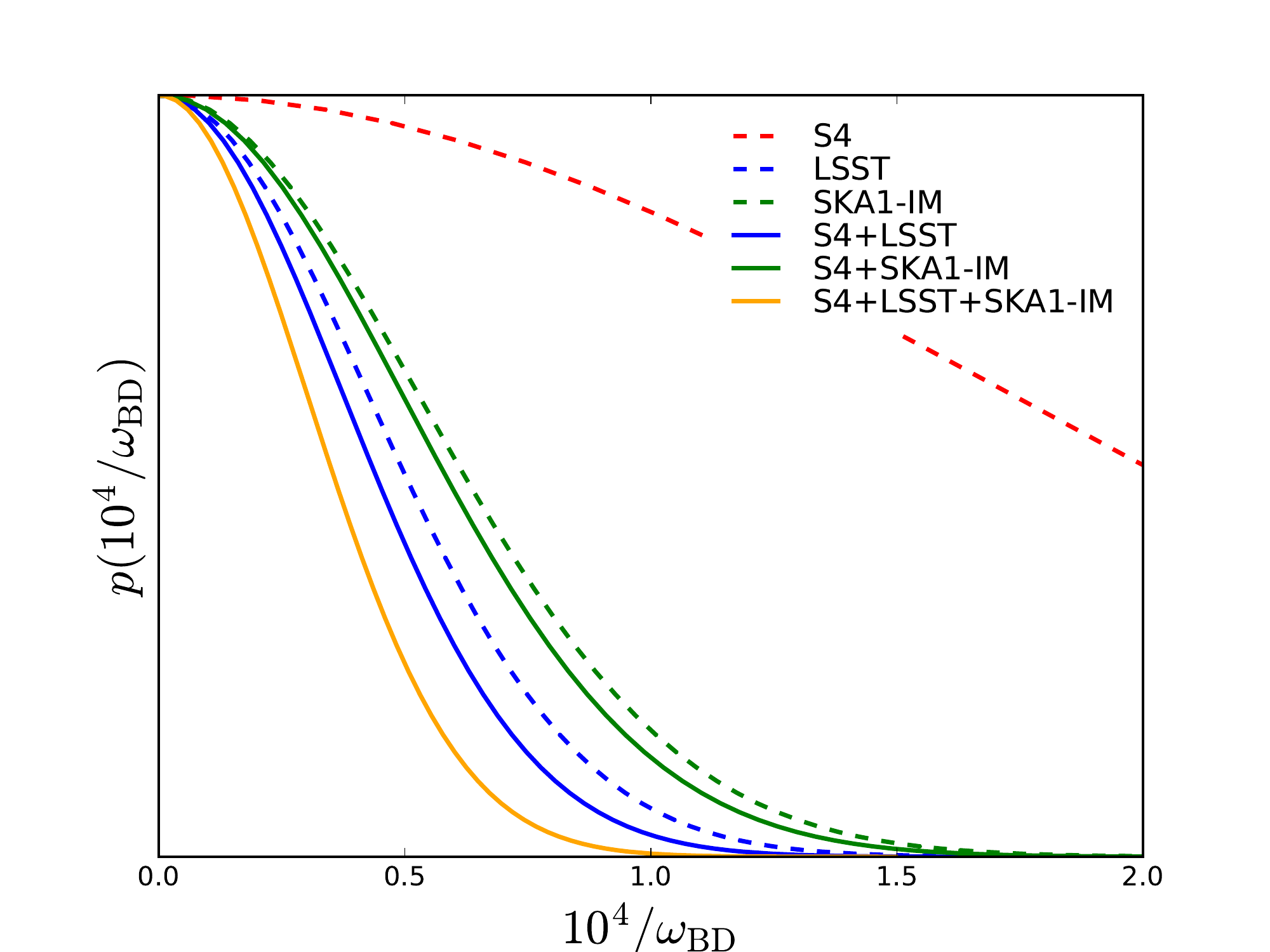}
	\caption{Posterior distribution of $\omega_{\rm BD}$ obtained combining a CMB-S4 survey with contemporary \textls[-15]{stage IV LSS experiments. Reprinted figure with permission from~the authors of \cite{alonso2017observational} by the American Physical~Society.}}\label{fig:cmbs4}
\end{figure}

An investigation on the role played by CMB with respect to LSS surveys was carried out in~\cite{alonso2017observational}. Together with more general scalar tensor theories, the~authors constrained the Jordan--Brans--Dicke theory obtaining bounds on the parameter $\omega_{\rm BD}$. In~Figure~\ref{fig:cmbs4}, we report the results found by the authors, which highlight how, even though CMB obtains constraints at redshifts where the modifications to gravity have no significant impact and it is therefore not competitive on its own, it can still be used to improve the constraints achieved with~LSS.

\unskip

\subsection{CMB-LSS Cross~Correlation}

While the improvement brought by the upcoming CMB and LSS surveys will be significant, a~crucial boost in the sensitivity on departures from GR could be achieved considering the cross-correlation between CMB and LSS observables. The~correlation between CMB temperature and galaxy positions allows to improve the sensitivity to the ISW effect, which would be instead severely limited by cosmic variance. This effect depends directly on the time derivatives of the Bardeen potentials $\Phi$ and $\Psi$ which can be changed significantly if one deviates from GR and, therefore, can potentially bring to strong constraints on such departures (see, e.g., in~\cite{Giannantonio:2009gi}). 

Such a strong synergy was explored for the combination of {\it Euclid} with present and future CMB experiments, i.e.,~{\it Planck}, SO and CMB-S4~\cite{Euclid:2021qvm}. This analysis was done on the $\gamma$ parameterization, thus in a particular framework where only the growth of structure is modified, while the lensing effect is unchanged with respect to~GR.

\begin{figure}[H]
	
	\includegraphics[width=0.85\columnwidth]{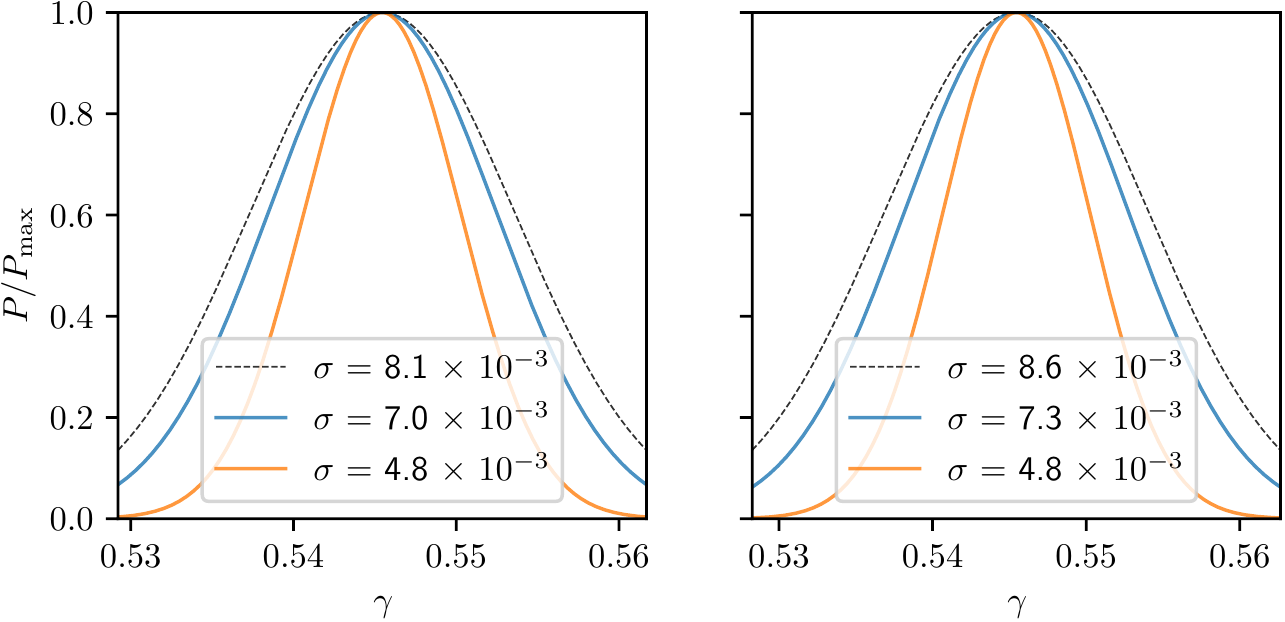}
	\caption{1D posterior distribution for the $\gamma$ parameter for {\it Euclid} alone (black), and~adding CMB lensing (blue) and all CMB observables (orange). The~CMB data here are taken to be SO-like. The~left panel refers to a flat $w_0w_a\gamma$CDM cosmology, while the right one drops the flatness assumption. This figure is taken from in~\cite{Euclid:2021qvm}.}\label{fig:xcmb}
\end{figure}

The authors of~\cite{Euclid:2021qvm} found that the inclusion of CMB data and its cross-correlation with {\it Euclid} observables will significantly improve the bound on the $\gamma$ parameter that determines the deviation from GR. In~Figure~\ref{fig:xcmb}, we show the results reported in~\cite{Euclid:2021qvm} for the bounds on $\gamma$; the posterior distributions shown highlight that while including only CMB lensing to {\it Euclid} does not significantly improve the constraints, adding the full CMB observables and their correlation with {\it Euclid} leads to a significant improvement both when assuming a flat Universe and when dropping the flatness~assumption.

\unskip

\subsection{New Probes of~Gravity}\label{sec:newprobes}

In the previous subsections, we focused on the expected cosmological tests of gravity that could be achieved thanks to upcoming CMB and LSS surveys. These constraints rely on improving the sensitivity with respect to current surveys, but~still rely on the same observational probes to constrain deviations from~GR.

However, recent observations have made available several new probes that could contribute in a significant way to the investigation of the theory of~gravity. 

After the first detection of a gravitational wave event~\cite{LIGOScientific:2016aoc}, significant interest has been devoted to this new window on cosmological evolution. Concerning the investigation of the gravity theory, the~observation of gravitational waves shows significant promise; as the existence of such a phenomenon is seen as one of the most striking validation of Einstein's theory, deviations from the expected behavior of such events would constituted a smoking gun for violations of GR. Indeed, one expects that if the gravity theory is different from GR, several effects could potentially be observed when detecting gravitational waves, e.g.,~a difference between the propagation speed of light ($c$) and gravitational waves ($c_T$), the~emission of additional polarizations and a modified damping of the waves amplitude~\cite{Ezquiaga:2018btd}.

The difference between $c$ and $c_T$ has been already tightly constrained thanks to a single GW event with an electromagnetic counterpart~\cite{LIGOScientific:2017vwq,Goldstein:2017mmi}. Such an event allowed to measure the speed of both propagations, constraining their relative difference to be smaller than $\approx$10$^{-15}$ \cite{LIGOScientific:2017zic}. Such a strong constraint has significant consequences for the study of deviations from GR, with~several possible alternative models being ruled out by this observation~\cite{Ezquiaga:2017ekz,creminelli2017dark}.

With the future increase in the number of observed events, as~the amplitude of the observed waves is proportional to the inverse of the luminosity distance $d_L(z)$, we expect to be able to obtain distance--redshift relation catalogs through GW observations, if~electromagnetic counterparts are available. Such catalogs have the potential to provide significant information on the theory of gravity. Indeed, in~the presence of an additional $(1-\delta(t))$ factor in the friction term of gravitational waves, with~$\delta=0$ corresponding to the GR limit, one expects to find a difference between the luminosity distances measured through GW ($d_L^{\rm GW}$) and electromagnetic probes ($d_L^{\rm EM}$) \cite{Belgacem:2017ihm,Belgacem:2018lbp}
\begin{equation}\label{eq:gwlum}
    \frac{d_L^{\rm GW}(z)}{d_L^{\rm EM}(z)} = \exp{\left[-\int_0^z{\frac{\delta(z')}{1+z'}dz'}\right]}\, .
\end{equation}

This possibility has been extensively explored in the literature, with~forecast for the possible deviations from GR obtained for future GW observatories such as LISA~\cite{2017arXiv170200786A} and the Einstein Telescope~\cite{Maggiore:2019uih}.

\begin{figure}[H]
	
	\includegraphics[width=0.85\columnwidth]{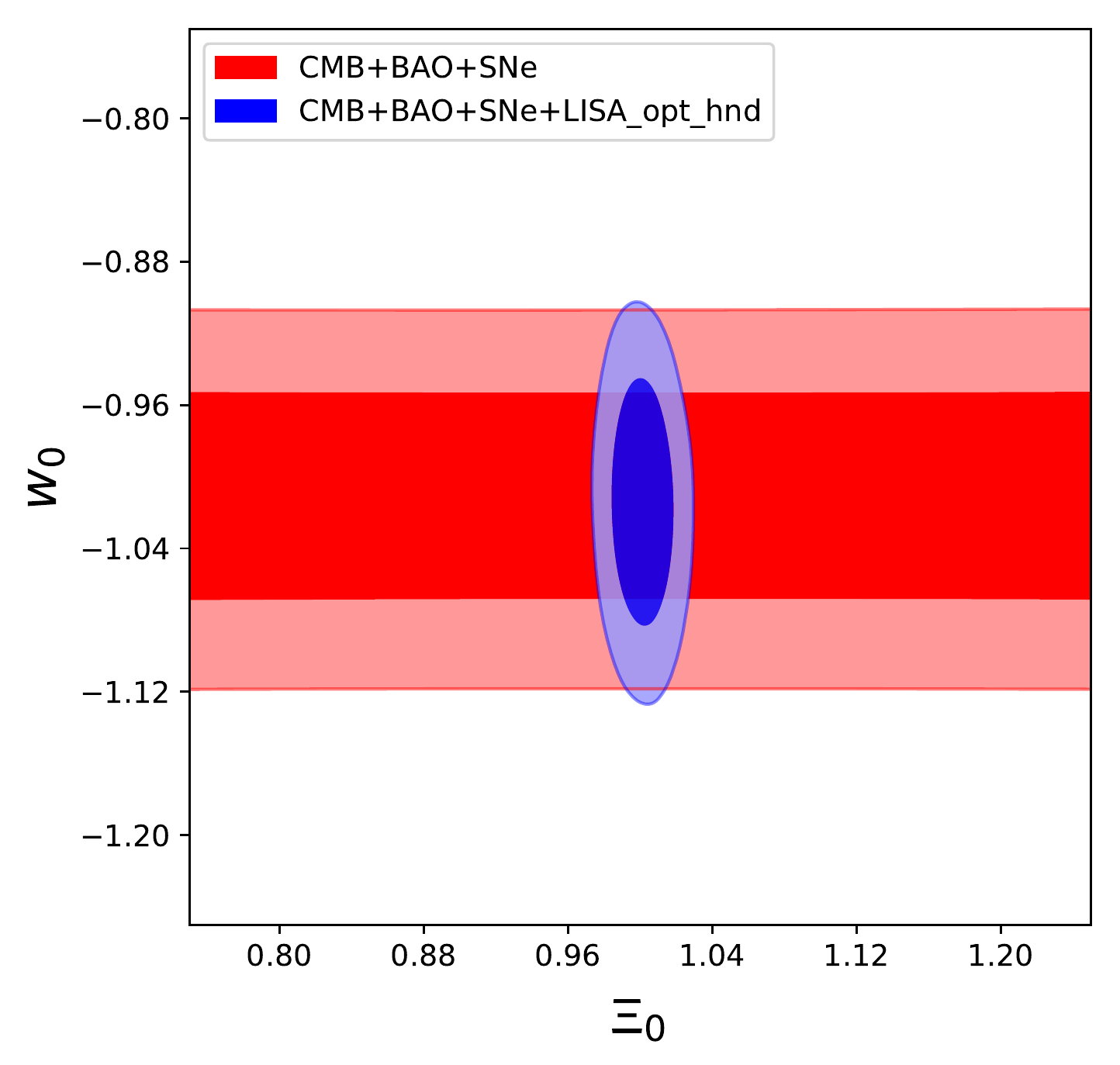}
	\caption{Constraints on $w_0$, parameterizing a constant DE equation of state, and~$\Xi_0$ obtained using CMB+BAO+SN (red) and when adding LISA forecast data to the previous combination (blue). This figure is taken from~\cite{LISACosmologyWorkingGroup:2019mwx}, where also different cases for the noise and for the seeds of the binaries have been investigated, and~reproduced by permission of IOP Publishing. All rights~reserved.}\label{fig:lisaMG}
\end{figure}

The LISA collaboration has indeed focused on this possible investigation with future observations, and~produced forecast on possible deviations from GR using a simulated catalog of massive binary black holes merger events~\cite{LISACosmologyWorkingGroup:2019mwx}. Here, the~authors parameterized Equation~\eqref{eq:gwlum} as
\begin{equation}\label{eq:gwlumpar}
    \frac{d_L^{\rm GW}(z)}{d_L^{\rm EM}(z)} = \Xi_0 + \frac{1-\Xi_0}{(1+z)^n}\, ,
\end{equation}
with $\Xi_0$ and $n$ free parameters. The~results of their analysis are shown in Figure~\ref{fig:lisaMG} which highlights how the constraining power of LISA can reach the level of $1\%$ on the parameter $\Xi_0$ which dictates the amplitude of deviations from the GR expected damping of GW amplitudes. Forecast constraints for the same parameterization of Equation~\eqref{eq:gwlumpar} were also obtained using a simulated Einstein Telescope catalog, finding a bound on $\Xi_0$ of $0.8\%$ \cite{Belgacem:2018lbp}. The~expected constraining power on deviations from GR of both LISA and Einstein Telescope are therefore very similar even though the two experiments explore complementary ranges in redshift; this highlights how the use of GW catalogs has the potential to test gravity over an extended redshift range and possibly rule out several of the alternative theories of gravity that are currently~available.

Despite the promise shown by these results, the~forecast reported above do not consider the impact of screening mechanisms on the constraints that can be achieved. Alternative theories of gravity need to be screened in the local environment to satisfy Solar System constraints. It could be expected that such screening mechanisms will be also at play in the local environment where the merger happens. In~such a case, both the observer and merger environments would be screened from MG effects. While most of the propagation area of a gravitational wave will not be screened, it has been shown that for screening mechanisms that fall into the chameleon class the deviation expected from the GR propagation only depends on the difference in the Planck mass at the observer and at the merger~\cite{Dalang:2019fma,Dalang:2019rke}. Such an effect would therefore completely remove any signature of departures from GR from the observations of GW luminosity~distances.

In addition to the possibility to directly test gravity with GW, it has been shown how GWs can be included in the set of observations used to probe the distance duality relation (DDR) \cite{Hogg:2020ktc}. The~DDR is a consequence of the assumptions at the foundation of our cosmological models, including the fact that it is based on a metric theory of gravity and that photons travel on null geodesic with their number conserved~\cite{EUCLID:2020syl}. Such a relation connects the luminosity and angular distance as
\begin{equation}
    d_L(z) = (1+z)^2d_A(z)\, .
\end{equation}

While deviations from GR are not the only possible responsible for a deviation from the DDR, indeed a detection of this would prompt a serious investigation on the nature of~gravity. 

\begin{figure}[H]
	
	\includegraphics[width=0.85\columnwidth]{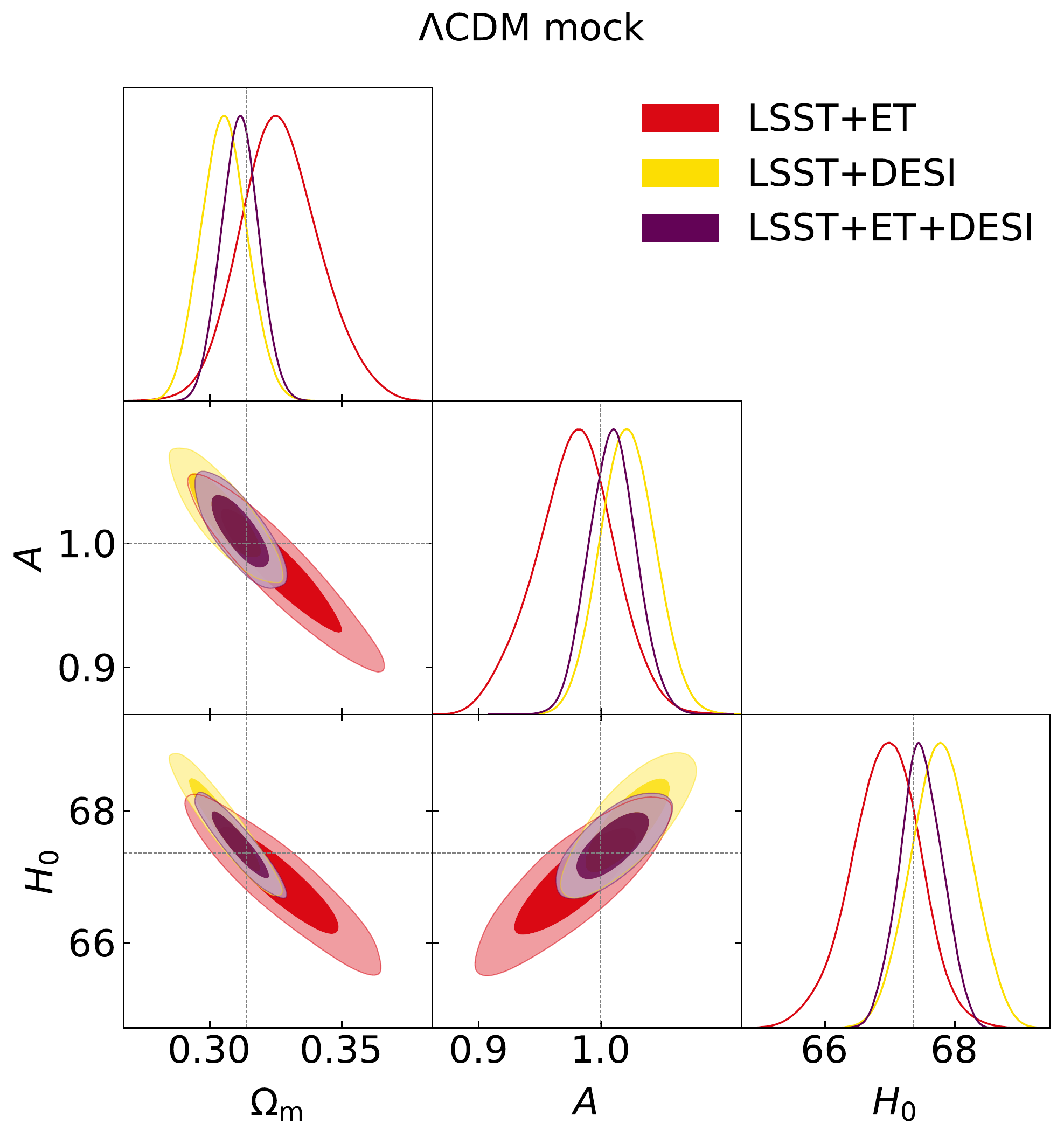}
	\caption{Constraints on cosmological parameters and on the DDR breaking parameter A. The~results are obtained for LSST+ET (red), LSST+DESI (yellow) and LSST+ET+DESI (purple). This figure is taken from in~\cite{Hogg:2020ktc}.}\label{fig:DDR}
\end{figure}

While the common observations to probe the DDR are supernovae (providing data on $d_L$) and BAO (measuring $d_A$), the~independent luminosity distance measure that can be obtained with GWs could be crucial for such an investigation. As~it does not rely on the observation of photons, a~measurement of $d_L^{\rm GW}$ could be used to break degeneracies between cosmological parameters and those ruling the deviation from the standard DDR. This possibility was investigated in~\cite{Hogg:2020ktc}, where the authors used a simulated catalog with Einstein Telescope specifications, alongside simulated data for SNIa and BAO. The~results found in their analysis are shown in Figure~\ref{fig:DDR} where it can be seen how GW data can contribute in tightening the constraints on the parameter $A$, which determines the amplitude of the deviation from DDR, with~the standard relation recovered for $A=1$.

We conclude this overview by mentioning another probe that is becoming increasingly important for cosmology: strong lensing time delays (SLTD). The~observation of SLTD has recently provided constraints on standard cosmological parameters, and~it has been used to obtain an independent measurement for $H_0$ with respect to the one inferred from CMB or obtain from the distance ladder (see, e.g.,~in \cite{Birrer:2020tax}). At~the same time, such an observable can be used to constrain deviations from GR; bounds on the parameterized post-Newtonian parameter $\gamma_{\rm PPN}$, which modifies the Weyl potential entering the lensing equations, have been obtained with current observations of strong lensing system~\cite{Jyoti:2019pez,Yang:2020eoh}, highlighting how future observations will be able to improve such bounds. Forecast on the constraining power of these observations have been done at the level of deviations from the standard $\Lambda$CDM background expansion~\cite{Shiralilou:2019div}, while a multi-messenger approach, studying the possible detection of strongly lensed GW events, could be used to constrain deviations from the GR limit of the $\Sigma$ function Equation~\eqref{eq:Sigma-def} at an $8\%$ level~\cite{Yang:2018bdf}.

\section{New~Challenges}\label{sec:newchallenges}

While the continuous improvements of observations will provide a great opportunity to test our gravity model, the~increased sensitivity and the possibility to access previously unavailable scales will also put to test our ability to obtain accurate theoretical predictions without introducing biases due to approximations or~assumptions. 

One very known example of this possible issue is the modeling of nonlinear scales. LSS surveys will provide data at scales where the linear perturbation theory used to obtain theoretical predictions in any cosmological model cannot be used, as~the amplitude of cosmological perturbations will be too high to use a linear approximation. If~the analysis one performs is limited to the standard $\Lambda$CDM model, this is not an issue; several N-body simulation performed for this model have allowed to obtain fitting functions that can complement our analytical theoretical predictions in the range of scales where these are not available. Codes like \texttt{Halofit} \cite{Takahashi:2012em} or \texttt{HMCode} \cite{Mead:2020vgs} have been integrated in the Boltzmann solvers \texttt{CAMB} and \texttt{CLASS} and allow to obtain predictions also at nonlinear~scales.

However, one cannot rely on these codes if the goal of the analysis is to test gravity; the N-body simulations used to obtain the fitting functions assume GR and $\Lambda$CDM, and~their use in different contexts can lead to significant biases in the parameter estimation pipeline, already when they are applied to very minimal extensions of the standard concordance model~\cite{Euclid:2020tff,Safi:2020ugb}.

One possible solution is to include in the analysis a theoretical error, modeled in such a way that it can account for the possible biases on the nonlinear theoretical predictions~\mbox{\cite{2013JCAP...01..026A,Sprenger:2018tdb,Knabenhans:2021huw}}. Such a method allows reducing the biases that incorrect modeling would propagate to the final estimate of the parameters, but~at the expense of larger uncertainties. Moreover, while this approach works for simple extensions of the $\Lambda$CDM, it might not be trivial to model the theoretical uncertainties when there is a more significant departure from the standard model, as~is the case for deviations from~GR.

Ideally, one would want to perform N-body simulation in all the models of interest, in~order to obtain fitting functions similar to those obtained in $\Lambda$CDM. This has been done for some specific models alternative to GR. 
Several interesting approaches have been taken in the literature such as \texttt{COLA} \cite{winther2017cola}, \texttt{Ramses} \cite{PhysRevD.89.084023}, \texttt{Ecosmog} \cite{li2012ecosmog},  $\phi$-\texttt{enics} (an interesting finite-element method approach that can capture the nonlinear evolution of a scalar field)~\cite{braden2021varphienics} and the simulation work on $f(R)$ theories by several groups~\cite{puchwein2013modified, reverberi2019frevolution, baldi2014cosmic, Valogiannis:2016ane}. However, applying such an approach for all available models, or~for model independent tests of gravity (performed through the $\mu$ and $\Sigma$ functions) is extremely difficult. In~\texttt{MG-evolution}~\cite{hassani2020n} the authors introduced the $\mu$-$\Sigma$ parameterization into a relativistic N-body code and managed to obtain predictions for those generalized models at nonlinear~scales. 

N-body simulations are computationally expensive and therefore only few realizations in a reduced parameter space are available for modified gravity models. Therefore, several approaches have appeared recently in order to reduce the required computational time and to allow for a more versatile evaluation of the observables. These include fitting formulae based on simulations~\cite{Winther:2019mus}, emulators for $f(R)$ and scalar-field theories~\cite{arnold2021forge, Ramachandra:2020lue, Mancini:2021lec} and hybrid approaches in which the halo model, perturbation theory and simulations are calibrated to create a model, such as \href{https://github.com/nebblu/ReACT}{\texttt{REACT}} (see~\cite{Bose2020, Cataneo:2018cic}). This code can compute predictions for $f(R)$ and other modified gravity models which are approximately 5\% accurate at scales $k \lesssim$ 5 h/Mpc.

While the linear scales are exempt from this specific issue, very large-scale pose different challenges to our modeling capabilities. LSS observables at the extremely large scales that will be probed by experiments such as {\it Euclid} and SKAO can indeed be modeled using the linear approach; however, additional contributions from relativistic effects become relevant at such scales~\cite{Yoo:2010ni,Challinor:2011bk,Bonvin:2011bg}, and, in~addition to this, the~approximations usually done to facilitate analytical computations (e.g., the Limber approximation~\cite{1953ApJ...117..134L,1954ApJ...119..655L,1992ApJ...388..272K}) are not valid anymore. Computing the exact analytical spectra is extremely expensive from the computational point of view, but~the use of approximated spectra can produce false detections of deviations from $\Lambda$CDM already when simple extensions are considered~\cite{Martinelli:2021ahc}. Such an effect would be particularly important for tests of the theory of gravity, as~deviations from GR are expected to produce significant signatures at these scales~\cite{2015ApJ...811..116B,Villa:2017yfg}. 

This issue has prompted several attempts to make the exact calculations less computationally expensive. For~instance, fast Fourier transform (FFT) or logarithmic FFT (FFTLog) methods can be exploited to accelerate the computation of the theoretical predictions~\cite{2017JCAP...11..054A,2017A&A...602A..72C,2018PhRvD..97b3504G}. Alternatives are represented by the use of emulators also in this case, as~these can allow to significantly speed up parameter estimation pipelines reducing the number of spectra for which the exact calculations need to be performed~\cite{Mancini:2021lec}, or~by the use of correction terms that allow to reduce the bias on parameter estimation~\cite{Martinelli:2021ahc}.

{In addition to this, exploiting the future data at extremely large-scale can raise the issue of validity of the quasi-static approximation, as~we discussed in} Section~\ref{sec:theory}, {further complicating the possibility of obtaining robust theoretical predictions when moving away from the GR paradigm.}

We conclude by briefly mentioning that the theoretical modeling of observables is not the only issue that the improvement of surveys will bring to light. Parameter estimation pipelines commonly assume a Gaussian likelihood when comparing theoretical predictions to data, an~assumption that might lead to inaccurate results given the extreme sensitivity that future data will have~\cite{Sellentin:2017koa}. Overcoming such an issue will require to either improve the theoretical modeling of the likelihood function or to resort to methods that completely eliminate the necessity of computing such a function (see, e.g.,~in \cite{Alsing:2019xrx,Jeffrey:2020xve}). While such an issue is not specific to cosmological tests of gravity, the~high sensitivity expected for future data requires the analysis to reach the best possible accuracy, in~order to fully exploit the available data to obtain robust and unbiased constraints on the theory of~gravity.

\section{Summary}\label{sec:summary}

The aim of this review was to give an outlook of the future possibilities to test the theory of gravity at cosmological scales. We briefly reviewed a subset of the available alternatives to the theory of general relativity, which is at the base of the current standard cosmological model $\Lambda$CDM. While it is indeed necessary to explore alternatives to GR from the theoretical point of view, here we focused mainly on how GR itself can be tested with cosmological observables, by~means of parameterized descriptions of departures from the expected cosmological evolution in GR. We briefly reviewed the most common parameterizations available, introducing free functions that allow cosmological perturbations to evolve differently with respect to the standard model, thus allowing to search from signatures of non-standard behavior in cosmological~observables.

We reviewed the state of the art of this investigation, highlighting how the most recent data, both from CMB and LSS, show a general agreement with GR, with~deviations from the standard paradigm that are not statistically significant ($\lesssim$2$\sigma$). Nevertheless, current data still leave open the possibility for a theory of gravity that does not coincide with~GR.

Upcoming surveys will therefore be crucial to further improve the bounds on departures from the standard paradigm, thanks to the increase in sensitivity and the possibility to probe a larger range of scales. We reviewed forecast results for this investigation of both upcoming CMB and LSS surveys, highlighting how the latter can in principle provide much stronger constraints with respect to the former. In~particular, we have shown how LSS surveys will be crucial to test gravity, thanks to their ability to probe the evolution of cosmological perturbations, through measurements of the matter distribution and growth rate, and~the lensing effect that matter has on the path of photons, through measurements of the shear of distant galaxies. The~two probes of GC and WL show a great complementarity and, if~combined, they would allow to tightly constrain for deviations from GR, both when analyzing specific classes of alternative models and when purely phenomenological parameterizations are~applied.

While future CMB surveys will have less constraining power on deviations from GR, their impact on tests of gravity will nevertheless be extremely important; we pointed out how the cross-correlation of CMB and LSS probes will allow to further improve the bounds on parameterized deviations from~GR.

Throughout this review, we have considered future CMB and LSS surveys as the main tools to test the theory of gravity, given the great expertise that the cosmological community has built on them through an extended period of time. However, we also highlighted how relatively new observations, such as strong lensing time delays and gravitational waves could also be used for this purpose. In~particular, we have shown how next-generation GW surveys will be able to tightly constrain deviations from the expected GR damping of the waves amplitude with distance, reaching an error of $\approx1\%$ on the parameter ruling the amplitude of such deviations. Moreover, we have also highlighted how the detection of strongly lensed gravitational wave could lead to independent constraints of the deviation from the lensing effect expected in GR, while the combination of GW observations with BAO and SNIa could improve the bounds on departure from the distance duality relation, which would signal that the fundamental assumptions at the base of our cosmological model should be~reviewed.

Finally, we focused our attention to the new challenges that we will face if we want to fully exploit this abundance of data that we will have available in the near future. We focused mainly on issues concerning the theoretical modeling of cosmological observables, pointing out that both very small and very large scale will put to test our ability to produce accurate theoretical predictions, showing how a failure from this point of view will make complicated to obtain robust tests of the theory of~gravity.

\vspace{6pt} 


\authorcontributions{Conceptualization, M.M. and S.C.; writing---original draft preparation, M.M. and S.C.; writing---review and editing, M.M. and S.C.; visualization, M.M.; supervision, M.M.; project administration, M.M. All authors have read and agreed to the published version of the~manuscript.}

\funding{MM has received the support of a fellowship from ``la Caixa'' Foundation (ID 100010434), with~fellowship code LCF/BQ/PI19/11690015, and~the support of the Centro de Excelencia Severo Ochoa Program SEV-2016-059.}

\acknowledgments{This paper is submitted to the Special Issue entitled ``Large Scale Structure of the Universe'', led by N. Frusciante and F. Pace , and~belongs to the section ``Cosmology''. We thank the authors of~\cite{Euclid:2019clj,Casas:2017eob,alonso2017observational,spurio2018testing,Euclid:2021qvm,LISACosmologyWorkingGroup:2019mwx} for granting us permission to use their figures in this~review.}

\conflictsofinterest{The authors declare no conflict of~interest.}

\begin{adjustwidth}{-4.6cm}{0cm}
\printendnotes[custom]
\end{adjustwidth}

\end{paracol}

\reftitle{References}


\end{document}